\documentclass[3p]{elsarticle}

\usepackage{fancyhdr}
\usepackage{hyperref}

\makeatletter
\def\ps@pprintTitle{%
 \let\@oddhead\@empty
 \let\@evenhead\@empty
 \def\@oddfoot{\footnotesize\itshape
\textcopyright 2021. Licensed under the CC-BY-NC-ND 4.0 user license \url{https://creativecommons.org/licenses/by-nc-nd/4.0/}}%
 \let\@evenfoot\@oddfoot}
\makeatother

\journal{Annual Reviews in Control}

\usepackage{amsmath} 
\usepackage{amssymb}
\usepackage{enumitem}  
\usepackage{pifont}
\usepackage{tikz}
\usepackage{tikz-cd}
\usetikzlibrary{shapes,arrows,fit,backgrounds,positioning}
\usetikzlibrary{decorations.pathmorphing} 
\usetikzlibrary{matrix} 
\usetikzlibrary{arrows} 
\usetikzlibrary{calc} 
\usepackage{verbatim}
\usepackage[table]{colortbl}

\newtheorem{definition}{Definition}[section]
\newdefinition{remark}{Remark}[section]
\newtheorem{assumption}{Assumption}[section]
\newtheorem{theorem}{Theorem}[section]

\newtheorem{property}{Property}[section]
\newtheorem{proposition}{Property}[section]
\newtheorem{corollary}{Corollary}[theorem]
\newdefinition{example}{Example}[section]

\newproof{proof}{Proof}
\newproof{pot}{Proof of Theorem \ref{thm2}}
\begin{document}

\begin{frontmatter}

\title{Koopman Operator Dynamical Models: Learning, Analysis and Control\tnoteref{t1}\tnoteref{funding}}

\tnotetext[t1]{\textbf{This is an authors' version of the work that is published in \textit{Annual Reviews in Control} journal. Changes were made to this version by the publisher prior to publication. The final version of record is available at \url{https://doi.org/10.1016/j.arcontrol.2021.09.002}.}}
\tnotetext[funding]{This work was supported by European Commission grant H2020-ICT-871295 ({"SeaClear" - SEarch, identificAtion and Collection of marine Litter with Autonomous Robots}).}
\author{Petar Bevanda\corref{cor2}}
\ead{petar.bevanda@tum.de}

\author{Stefan Sosnowski}
\ead{sosnowski@tum.de}

\author{Sandra Hirche}
\ead{hirche@tum.de}

\address{Chair of Information-oriented Control, Department of Electrical and Computer Engineering, Technical University of Munich, D-80333 Munich, Germany}

\cortext[cor2]{Corresponding author}

\begin{abstract}
The Koopman operator allows for handling nonlinear systems through a globally linear representation. In general, the operator is infinite-dimensional - necessitating finite approximations - for which there is no overarching framework. Although there are principled ways of learning such finite approximations, they are in many instances overlooked in favor of, often ill-posed and unstructured methods. Also, Koopman operator theory has long-standing connections to known system-theoretic and dynamical system notions that are not universally recognized. Given the former and latter realities, this work aims to bridge the gap between various concepts regarding both theory and tractable realizations.
Firstly, we review data-driven representations (both unstructured and structured) for Koopman operator dynamical models, categorizing various existing methodologies and highlighting their differences. Furthermore, we provide concise insight into the paradigm's relation to system-theoretic notions and analyze the prospect of using the paradigm for modeling control systems. 
Additionally, we outline the current challenges and comment on future perspectives. 

\end{abstract}


\end{frontmatter}

\section{INTRODUCTION}\label{secI}
	Traditionally, systems are represented in the immediate state space, concerned with ``dynamics of states''. 
	Although such representations enjoy incredible success, they reach limits when it comes to efficient prediction, analysis and optimization-based control. The aforementioned motivated an increased interest in a system representation inspired by Koopman operator theory, where a finite-dimensional nonlinear system is replaced by an infinite-dimensional linear one. The paradigm is named after B.O. Koopman - the author of the seminal work \cite{Koopman1931} on transformations of Hamiltonian systems in Hilbert space. The operator's existence had an important role in proving the Mean Ergodic Theorem of Neumann \cite{Neumann263} and Birkhoff \cite{Birkhoff1931}. The first consideration of the theory outside the measure-preserving context came through the works of Mezi\'{c} et al. \cite{Mezic2004,Mezic2005} which included ideas on finite-dimensional representations of the Koopman operator using eigenfunctions.

	For complex systems, models coming from first-principles often times are not available or do not fully resemble the true system due to unmodeled phenomena. The complexity of systems we encounter prompts a shift from classical parametric techniques in favor of more flexible machine learning techniques (e.g. neural networks or Gaussian processes) for prediction \cite{Nelles2001,Kocijan2005}, model-based control \cite{Umlauft2018, BECKERS2019390} and analysis \cite{Berkenkamp2016a,Fisac2018a,Lederer2019a}.
	The aforementioned learning approaches are, however, limited with regards to efficient prediction, analysis and control, due to  modeling the ``dynamics of states". Nevertheless, Koopman operator realizations are inherently data-driven as one needs to discover suitable coordinates that represent the operator - leading to globally linear models of nonlinear systems.
	As the modeling paradigm dictates the challenges in dealing with a system, trading infinite-dimensionality for linearity allows us to employ efficient linear techniques for nonlinear systems. The former leads to more challenging identification but efficacious prediction, analysis and control.
	
	 Representations via the Koopman paradigm generalize the notion of mode analysis from linear to nonlinear systems allowing for amendable relevance determination of the constituents of the full dynamics as for thermal analysis of buildings \cite{Georgescu2012}. Furthermore, the spectrum of the Koopman operator allows one to decompose the nonlinear system into different dynamic regimes (fast-slow dynamics) while linearly evolving coordinates contain intrinsic information relevant for analysis i.e. regions of attraction \cite{Mauroy2013b}. Due to the linear dynamics of the new coordinates, prediction hardships through numerical integration and non-convexity in optimization of ``dynamics of states'' have the potential to be alleviated as for model predictive control \cite{Korda2020b,Igarashi2020}.
	Moreover, one can argue that the {Koopman operator paradigm} delivers a global instead of a point-wise system description as one iteration of the Koopman operator acting on an observable is equivalent to an iteration along all of the trajectories of the system and it is not to be confused with a local linearization around a working point.
	
	Given the Koopman operator is generally infinite-dimensional, it necessitates finite realizations - a task for which there is no overarching, general framework. The non-triviality of finite representations lead to many data-driven approaches with various trains of thought, facilitating different properties of the Koopman operator paradigm. Regarding the plethora of data-driven frameworks looking to deliver Koopman operator dynamical models, this article - to the best of the authors' knowledge - is the first work reviewing the aforementioned in a systematic manner while giving theoretical insight.

A related review of the paradigm by Budi\v{s}i\'{c} et al. \cite{Budisic2012a} gives a theoretical baseline with application examples but with the data-driven methodology limited to dynamic mode decomposition (DMD). We address the abundance of novel data-driven methods that arose from and beside DMD for discovering Koopman operator dynamical models in addition to siding with a more focused system and control-theoretic perspective.
Also, compared to work of Kaiser et al. \cite{Kaiser2020}, we take a rigorous approach focused on the Koopman operator paradigm instead of an high-level overview of data-driven transfer operators. By taking such an approach, the aim is to deliver a holistic and methodical backbone of Koopman operator-based dynamical models - from surveying the data-driven representations, to system-theoretic connections and control.

The article is structured as follows. After the preliminaries on Koopman operator theory in Section \ref{secII}, the data-driven methods for Koopman-related model representation are presented in Section \ref{secIII}. An overview of system-theoretic analysis via Koopman operator theory follows in Section \ref{secIV}. Furthermore, Section \ref{secV} analyzes the inclusion of control into the representations and surveys their applications found in the literature. Before the final outline concluding Section \ref{secVII}, we comment on future perspectives of Koopman-based methods.

\section*{Notation}\label{notat}
Lower/upper case bold symbols $\boldsymbol{x}$/$\boldsymbol{X}$ denote vectors/matrices.
Symbols $\mathbb{N}/\mathbb{ R }/\mathbb{C}$ denote sets of natural/real/complex numbers while $\mathbb{N}_{0}$ denotes all natural numbers with zero and $\mathbb{R}_{+,0}/\mathbb{R}_{+}$ all positive reals with/without zero. 
The continuous/discrete-time dependence is denoted as $y^{t}(\cdot)$ and $y_{k}$, respectively with $t / k\in \mathbb{R}_{+,0}/ \mathbb{N}_0$ while $\dot{(\cdot)}:=d(\cdot)/dt$, for brevity. 
Function spaces with a specific integrability/smoothness order are denoted as $L^{}$/$C^{}$ with the order (class) specified in their exponent.
The braces $\left\langle \cdot , \cdot \right\rangle$ denote the inner product while $\|\cdot\|$ the Euclidean norm.
A flow induced by a vector field $\dot{x}=f(x)$ is denoted as $F^{t}(x)$ with its associated family of composition (Koopman) operators $\{\mathcal{K}^{t}_{f}\}_{t \in \mathbb{R}_{+,0}}$. A map i.e. $x \mapsto F(x)$ has its associated family of composition operator iterates denoted as $\{\mathcal{K}^n_F\}_{n \in \mathbb{N}_0}$.

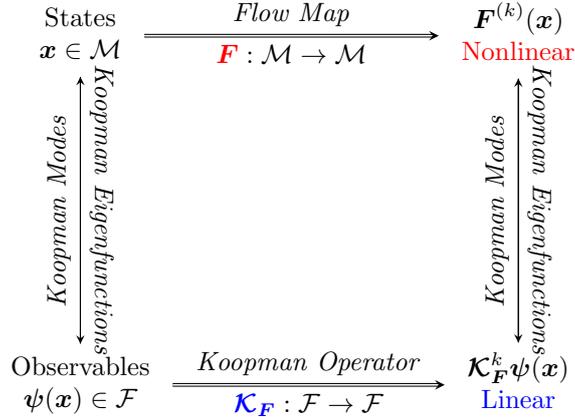
\begin{figure}[ht!]
	\centering
	\begin{tikzpicture}
	\matrix (m) [matrix of math nodes,row sep=10em,column sep=10em,minimum width=3em]
	{
		\begin{array}{c}
		\text{States}\\ \boldsymbol{x} \in \mathcal{M}
		\end{array} & \begin{array}{c}
		\boldsymbol{F}^{(k)}(\boldsymbol{x}) \\ {\color{red}\text{Nonlinear}}
		\end{array}  \\
		\begin{array}{c}
		\text{Observables}\\ \boldsymbol{{\psi}}(\boldsymbol{x}) \in \mathcal{F}
		\end{array} & \begin{array}{c}
		\boldsymbol{\mathcal{K}}^{k}_{\boldsymbol{F}}\boldsymbol{{\psi}}(\boldsymbol{x}) \\ {\color{blue}\text{Linear}}
		\end{array} \\};
	\path[-stealth]
	(m-1-1) edge node [above, rotate=90] {\textit{Koopman Modes}} (m-2-1)
	(m-2-1) edge node [above, rotate=270] {\textit{Koopman Eigenfunctions}} (m-1-1)
	(m-1-1) edge [double] node [below] {$ {\color{red}\boldsymbol{{F}}}: \mathcal{M} \rightarrow \mathcal{M}$} node [above] {\textit{Flow Map}} (m-1-2)
	(m-2-1) edge[double] node [below] {$ {\color{blue}\boldsymbol{\mathcal{K}}_{\boldsymbol{F}}}: \mathcal{F} \rightarrow \mathcal{F}$} node [above] {\textit{Koopman Operator}} (m-2-2)
	(m-2-2) edge node [above, rotate=90] {\textit{Koopman Modes}} (m-1-2)
	(m-1-2) edge node [above, rotate=270] {\textit{Koopman Eigenfunctions}} (m-2-2);
	\end{tikzpicture}
	\caption{Diagram of the Koopman operator system modeling concept ($\boldsymbol{F}^{(i)}=\boldsymbol{F} \circ \ldots \circ \boldsymbol{F}$ $i$-time iteration). The symbol $\boldsymbol{\mathcal{K}}=\operatorname{diag}\{\mathcal{K},\dots,\mathcal{K}\}$ is there for element-wise multiplicative application to vector-valued observable $\boldsymbol{{\psi}}$.} 
	\label{quantities}
\end{figure}

{\section{Koopman operator theory}\label{secII}
Let us commence with the basic assumptions and definitions required for the introduction of the {Koopman operator paradigm}.
\begin{assumption}\label{assSysF1}
	Consider a continuous-time dynamical system
	\begin{equation}\label{minSys}
	\dot{\boldsymbol{x}}=\boldsymbol{f}(\boldsymbol{x}),
	\end{equation}
	and a scalar \emph{observable} function ${\psi}: \mathcal{M} \mapsto \mathbb{C}$
	\begin{equation}\label{obs}
	y={\psi}(\boldsymbol{x}),
	\end{equation}
	on a {smooth} $n$-dimensional  manifold $\mathcal{M}$, $\boldsymbol{x} \in \mathcal{M}$, $t \in \mathbb{R}_{+,0}$ with a smooth and Lipschitz {flow} $\boldsymbol{F}^{t}: \mathcal{M} \mapsto \mathcal{M}$ \cite{Lasota1994}:
	\begin{equation}\label{flow}
	\boldsymbol{x}(t_0) \equiv \boldsymbol{x}_0, \quad \boldsymbol{F}^{t}(\boldsymbol{x}_0):=\boldsymbol{x}_0+\int_{t_{0}}^{t_{0}+t} \boldsymbol{f}(\boldsymbol{x}(\tau)) d \tau,
	\end{equation}
	with (\ref{flow}) forward-complete \cite{Bittracher2015}, i.e. $\boldsymbol{F}^{t}(\boldsymbol{x})$ has a unique solution on $[0,+\infty)$ from the initial condition $\boldsymbol{x}$ at $t = 0$ \cite{Angeli1999a}.
\end{assumption}
The system class satisfying the above assumption is quite substantial and it includes many physical systems \cite{Krstic2009}.

In most cases the iteration of a fixed time-$t$ flow map is considered to represent the dynamical evolution - where the notion of \textit{iteration} represents consecutive compositions of a map with itself.	
In Figure \ref{ogNLSysEvo} we can recognize the traditional study of nonlinear system in terms of orbits of points $\boldsymbol{x}$ on a domain $\mathcal{M}$ as \textit{dynamics of states} \cite{Budisic2012a}.
In comparison, Figure \ref{EFsliceEVO} shows that the Koopman operator entails \emph{how hypersurfaces (observables) over the state space propagate over time}. An observable can be any kind of a measurement of the system, e.g. a sensory measurement.
In the following, we formally define the operator.
\begin{definition}[Koopman Operator] \label{defKoop}
	For dynamical systems satisfying Assumption \ref{assSysF1}, the semigroup of Koopman operators $\{{\mathcal{K}}^{t}\}_{t \in \mathbb{R}_{+,0}}: \mathcal{F} \mapsto \mathcal{F}$ acts on scalar observable functions ${\psi}: \mathcal{M} \mapsto \mathbb{C}$ by composition with the flow semigroup $\{\boldsymbol{F}^{t}\}_{t \in \mathbb{ R }_{0,+}}$ of the vector field $\boldsymbol{f}$
	\begin{equation}\label{KoopEvo}
	{\mathcal{K}^{t}_{\boldsymbol{f}}} {{\psi}} ={{\psi}}\circ{\boldsymbol{F}^{t}},
	\end{equation}
	on the state space $\mathcal{M}$.
	When dealing with a map (discrete-time system) we have
	\begin{equation}\label{KoopEvoDT}
	{\mathcal{K}_{\boldsymbol{F}}} {{\psi}} ={{\psi}}\circ{\boldsymbol{F}},
	\end{equation}
	{where $\circ$ represents function composition.} Given that, the Koopman operator is also referred to as the \emph{composition operator}. In the remainder of this work we use these terms interchangeably.
\end{definition}

\begin{property}[Linearity of $\mathcal{K}$-operator] \label{linK}
	Consider the Koopman operator $\mathcal{K}_{\boldsymbol{F}}$ and two observables $g_1$, $g_2 \in \mathcal{F}$ and a scalar $\alpha \in \mathbb{R}$. 
	Using (\ref{KoopEvoDT}) it follows that
	\begin{equation}
	\begin{aligned}
	\mathcal{K}_{\boldsymbol{F}}(\alpha g_1+g_2)& = (\alpha g_1+g_2)\circ \boldsymbol{F} \\
	& =\alpha g_1 \circ \boldsymbol{F}+g_2 \circ \boldsymbol{F} \\ &=\alpha \mathcal{K}_{\boldsymbol{F}} g_1 + \mathcal{K}_{\boldsymbol{F}} g_2,
	\end{aligned}
	\end{equation}
	showing the linearity of the operator.
\end{property}
\begin{remark}
	Compared to compositions of nonlinear maps, the Koopman operator acts \emph{multiplicative in time} - due to Property \ref{linK} - as shown in Figure \ref{quantities}.
\end{remark}
}

\begin{assumption}\label{ass:3}
	There is a generator $\mathcal{G}_\mathcal{K}: D(\mathcal{G}_\mathcal{K}) \rightarrow \mathcal{F}$, $D$ being the domain of the generator and $\mathcal{F}$ the (Banach) space of observables.
\end{assumption}

\begin{definition}[Infinitesimal generator]\label{generator}
	The operator ${\mathcal{G}_{\mathcal{K}}}$, is the infinitesimal generator of the time-$t$ indexed semigroup of Koopman operators $\{{\mathcal{K}}^{t}\}_{t \in \mathbb{R}_{+,0}}$ 
	\begin{equation}
	\mathcal{G}_{{\mathcal{K}}} {\psi}=\lim _{t \rightarrow 0^{+}} \frac{{\mathcal{K}}^{t} {\psi}-{\psi}}{t} = \frac{d}{dt}{\psi},
	\end{equation}
	\cite[Section 3.3]{Lasota1994}.
\end{definition}

For a bounded generator $\mathcal{G}_{\mathcal{ K }}$, one obtains the time-$t$ semigroup of Koopman operators as $\mathcal{ K }^{t} = \operatorname{exp}(\mathcal{G}_{\mathcal{K}} t )$.
Analogous to the Koopman operator we use the same paradigm to get the linear evolution of the observables for the Koopman operator generator.
{Many nonlinear physical systems are continuous-time in nature and thus hard to meaningfully discretize - making a good use-case for linear-in-time Koopman generator representations to obtain exact discretization of the system's modes.}

\subsection{Koopman operator and generator eigendecomposition}

To replace a nonlinear system with a linear one, the linearity from Property \ref{linK} is only a helper tool and the key is the existence of linearly evolving coordinates - the Koopman operator \emph{eigenfunctions}.
\begin{definition}[Koopman eigenfunction]\label{eigF}
	An observable $\phi \in \mathcal{F}$ is an \textit{eigenfunction} if it satisfies
	\begin{equation}\label{KGEVE}
	[\mathcal{G}_{\mathcal{K}} \phi](\boldsymbol{x}) = \dot{\phi} \left(\boldsymbol{x}\right) = s \phi (\boldsymbol{x}),
	\end{equation}
	associated with the \textit{eigenvalue} $s \in \mathbb{C}$.
\end{definition}

\begin{property}[Linear coordinates]\label{linCoor}
	Since $\mathcal{G}_{\mathcal{K}}$ is the infinitesimal generator of the semigroup of Koopman operators $\{{\mathcal{K}}^{t}\}_{t \in \mathbb{R}_{+,0}}$, the following is also satisfied for  $\lambda^t = \operatorname{exp}({s t})$
	\begin{equation}\label{eq:eigF}
	[\mathcal{K}^{t}_{\boldsymbol{f}} \phi](\boldsymbol{x}) = \phi\left( \boldsymbol{F}^{t} (\boldsymbol{x}) \right) = \lambda^t \phi \left(\boldsymbol{x} \right),
	\end{equation}
	along the vector field's flow.
\end{property}

For the discrete-time case, the Koopman operator $\mathcal{K}_{\boldsymbol{F}}$ composes an observable with the map $\boldsymbol{x} \mapsto \boldsymbol{F}(\boldsymbol{x})$ not parameterized by time (as opposed to a flow of an ODE) and in that case delivers a simpler and arguably more intuitive eigenvalue equation
\begin{equation}\label{eq:EVE}
[\mathcal{K}_{\boldsymbol{F}} \phi](\boldsymbol{x}) = {\phi}\left(\boldsymbol{F}(\boldsymbol{x}) \right) = \lambda \phi \left(  \boldsymbol{x} \right),
\end{equation}
with $\lambda=\operatorname{exp} \left(s t_s\right)$, where $t_s$ is the sampling time discrete-time dynamical system.

To exploit the property of eigenfunctions being linear coordinates, we need to be able to express the evolution of an arbitrary observable using them. With that we can claim that we captured the effect of the Koopman operator which is crucial for finite approximations that are presented in Section \ref{secIII}.

	\begin{property}[Spectral decomposition]\label{OFEF}
		We can write an observable via the eigenfunctions of the Koopman operator
		\begin{equation}\label{decomp}
		{\psi} (\boldsymbol{x}) = \sum_{j=0}^{\infty} v_j (\psi) \phi_j({\boldsymbol{x}}) ,
		\end{equation}
		where the coefficients ${v}_j$ are Koopman operator \emph{modes} of the corresponding \emph{eigenfunctions} $\phi_j$.
		Note that associating modes to eigenvalues leads to a loss of generality \cite[Remark 10]{Budisic2012}.
	\end{property}
	
	Obviously, the infinite-dimensionality of the Koopman operator is due to the infinite-sum in (\ref{decomp}) that should not be confused with infinite-series expansions i.e. Taylor/Carleman \cite{Carleman1932} that do not necessarily provide dynamically closed (invariant) coordinates nor preserve global dynamical properties \cite{Gu2011}.
	The Koopman operator acts on an observable {function} as follows
	\begin{equation}\label{evoOF}
	\mathcal{K}^{t} {\psi} =\sum_{j=1}^{\infty} {v}_{j}(\psi)\left(\mathcal{K}^{t} \phi_{j}\right)=\sum_{j=1}^{\infty} {v}_{j}(\psi) \lambda^t_{j} \phi_{j},
	\end{equation}
	where $\lambda^t_j = \operatorname{exp}({s_j t})$ s.t. $s_j=\gamma_j + i\omega_j$ with eigen-decay/growth $\gamma_j$ and {eigenfrequencies} $\omega_j$ \cite{Mezic2013}. The above introduced decomposition is clearly amendable for linear prediction, however it is only a part of the Koopman operator's spectrum. Although the most successful dynamical models in the field focus on the point spectrum, it is only a part of the full spectrum. In the following we extend to the full spectrum for completeness.

	\begin{remark}[Full spectrum]\label{rmk:fullSPECT}
		In general, the spectral decomposition can include the continuous spectrum as well \cite{Mezic2005}. Given the system has an attractor $\mathcal{A}$ with a preserved measure $\mu(\mathcal{A})$ \cite{Arbabi2017} we write the Koopman operator spectrum as
		\begin{equation}\label{eq:fullSpect}
		\mathcal{K}^{t} {\psi}(\boldsymbol{x}) = \underbrace{\sum_{j=1}^{\infty} {v}_{j}(\psi) \lambda^t_j \phi_{j}(\boldsymbol{x})}_{\text{point}} +\\
		\underbrace{\int_{0}^{2\pi} \kappa^t E(d\kappa) \psi(\boldsymbol{x})}_{\text{continuous}},
		\end{equation}
		where $\kappa^t = \operatorname{exp}(i\theta t)$ is the integrated function over frequency $\theta$ and $E(\kappa)$ the projection-valued measure. While the point spectrum extracts transient - $\operatorname{Re}(\lambda^t_j)$ and oscillatory components of dynamics - $\operatorname{Im}(\lambda^t_j)$ (quasi-periodic \cite{Takeishi2019}), the continuous part encodes the ``chaotic'' component of the dynamics correspond to the on-attractor dynamics. The latter can also be seen as the extension of the notion of the point spectrum, where eigenfunctions are replaced by eigenmeasures \cite{Mezic2019a}.
	\end{remark}
	For a connection of the operator's spectrum and it's resolvent generalizing Laplace-domain theory to nonlinear dynamics, we point to Susuki et al. \cite{Susuki2020a}. 
	
\begin{property}[Koopman eigenfunction {group}]\label{productEFs}
Under the assumption that the space $\mathcal{F}$ is chosen to be a Banach algebra (e.g. $C^1(\mathcal{M})$) the set of eigenfunctions forms an Abelian semigroup under point-wise products of functions \cite{Budisic2012}. Thus, products of eigenfunctions are, again, eigenfunctions - if $\phi_{1}, \phi_{2} \in \mathcal{F}$ are eigenfunctions of the composition operator $\mathcal{K}_{\boldsymbol{F}}$ with eigenvalues $\lambda_{1}$ and $\lambda_{2},$ then $\phi_{1} \phi_{2}$ is an eigenfunction of $\mathcal{K}_{\boldsymbol{F}}$ with eigenvalue $\lambda_{1} \lambda_{2}$.
\end{property}

{ 	
	\begin{definition}[Principle eigenpairs]\label{princEP}
		Consider $\{E^m\}_{m \in \mathbb{N}}$ to be the eigenpair-semigroup of Koopman operator $\mathcal{K}_{\boldsymbol{F}}$ with its minimal generator $\mathcal{G}_{E}$:
		\begin{equation}\label{PeP}
		\{E^m\}=\left\{\left(\prod_{i=1}^{m} \lambda_{i}^{n_{i}}, \prod_{i=1}^{m} \phi_{i}^{n_{i}}\right) \mid\left(\lambda_{i}, \phi_{i}\right) \subset \mathcal{G}_{E} \right\},
		\end{equation}
		where $m, n_{i} \in \mathbb{N}$ \cite{Mohr2014a}. Then, the elements of  $\mathcal{G}_{E}$ are \emph{principle} eigenvalues with corresponding eigenfunctions of $\mathcal{K}_{\boldsymbol{F}}$ in $\{E^m\}_{m \in \mathbb{N}}$.
	\end{definition}
	{
	\begin{remark}
		Less formally, \emph{principle eigenpairs} form the minimal set of Koopman operator eigenpairs that can then be used to construct all other eigenpairs (\ref{PeP}). This notion is first considered for linear systems in \cite{Mohr2014a} and extended to nonlinear systems to some extent in \cite{Kvalheim2019}. 
		Our Definition \ref{princEP} helps the simplicity of exposition (although it originates from linear systems \cite{Mohr2014a}). However, the repercussions of such a definition are not rigorously studied with the exception of Bollt et al. \cite{Bollt2021} under the name ``primary" eigenfunctions.
	\end{remark}
}

	To give perspective and embed some of the dominant ideas when it comes to learning Koopman operator-based representations (Section \ref{secIII}), we present the following nonlinear discrete-time dynamical system that admits an exact finite-dimensional linear representations of the Koopman operator action.
	\begin{example}[Motivating example]\label{ex:motiv}
		Consider the map $\boldsymbol{F}: \mathbb{ R }^2 \mapsto \mathbb{ R }^2$ inspired by Tu et al. \cite{Tu2014}:
		
		\begin{equation}\label{SMdt}
		\boldsymbol{x} \mapsto \boldsymbol{F}(\boldsymbol{x}) = \left[\begin{array}{c}
		a x_{1} \\
		b x_{2}+\left(b-a^{2}\right) x_{1}^{2}
		\end{array}\right] ,
		\end{equation}
		{where $a$ and $b$ $\in$ $[0,1]$. The system has a stable equilibrium at the origin and invariant manifolds given by $x_{1}=0$ and $x_{2}=-x_{1}^{2}$.} 
		
		\begin{enumerate}[label=(\alph*)]
			\item \emph{Eigenfunction coordinates:}\label{ex:motivEF}
			The associated \emph{principle} eigenvalue-eigenfunction pairs $(\lambda,\phi(\boldsymbol{x}))$ of the system are $(a,{x}_1)$ and $(b,{x}_2+{x}_1^2)$.
			Utilizing the closedness of eigenfunctions under point-wise products from Property \ref{productEFs} we can obtain more eigenfunction-eigenvalue pairs as powers of the principle ones i.e. $(a^2,{x}^2_1)$ - a product of $(a,{x}_1)$ with itself. Utilizing the newly obtained pair, the system from (\ref{SMdt}) can be \emph{lifted} into new linear-in-time coordinates $\boldsymbol{\phi}(\boldsymbol{x})=[x_1, x_2+ x_1^2, x_1^2]^{\top}$:
			
			\begin{equation}\label{EFmap}	
			\boldsymbol{\phi}(\boldsymbol{x}) \mapsto \underbrace{\left[\begin{array}{ccc}
				a & 0 & 0 \\
				0 & b & 0 \\
				0 & 0 & a^2 
				\end{array}\right]}_{\boldsymbol{{ \Lambda }}} \boldsymbol{\phi}(\boldsymbol{x}).
			\end{equation}
			
			The result of (\ref{EFmap}) can be then be projected onto the respective vector of \emph{output functions} e.g. system states $\boldsymbol{\psi}(\boldsymbol{ x})=[x_1,x_2]^{\top}$ using Property \ref{OFEF} via Koopman operator modes $\boldsymbol{v}_1=[1, 0]^{\top}$, $\boldsymbol{v}_2=[0, 1]^{\top}$ and $\boldsymbol{v}_3=[0, -1]^{\top}$
			\begin{equation}\label{eq:koopEF}
			[\boldsymbol{\mathcal{K}}_{\boldsymbol{F}}](\boldsymbol{x})	=\boldsymbol{F}(\boldsymbol{x}) = {\boldsymbol{V}}\boldsymbol{\Lambda}\boldsymbol{\phi}(\boldsymbol{ x})
			\end{equation}
			with $\boldsymbol{V} = [{\boldsymbol{v}_1 \quad \boldsymbol{v}_2 \quad \boldsymbol{v}_3}]$ and offer an equivalent representation for the system (\ref{SMdt}). {A derivation of the solution can be found in the subsequent section, see also Remark \ref{origCONJ}.}
			With that, the system's propagation does not necessitate the composition of nonlinear maps but only an initial nonlinear transformation after which the evolution is linear
			\begin{equation}\label{iterEFs}
			\boldsymbol{F}^{(k)}(\boldsymbol{x}) = {\boldsymbol{V}}\boldsymbol{\Lambda}^{k}\boldsymbol{\phi}(\boldsymbol{ x}) 
			\end{equation}
			where $k$ compositions of nonlinear functions are replaced with \emph{lifting} $\boldsymbol{\phi}(\cdot)$ and subsequent linear evolution 
			through the \emph{composition} operator paradigm.
			
			\item \emph{Observables as coordinates:}\label{ex:motivOF}
			Interestingly, the dominant idea in the Koopman operator paradigm is to find ``good" coordinates what are not necessarily eigenfunctions but lie in the span of eigenfunctions. For the same system (\ref{SMdt}) the choice of coordinates  $\boldsymbol{\psi}(\boldsymbol{x})=[x_1, x_2, x_1^2]^{\top}$ still leads to the following linear mapping
			\begin{equation}\label{OBmap}	
			\boldsymbol{\psi}(\boldsymbol{x}) \mapsto \underbrace{\left[\begin{array}{ccc}
				a & 0 & 0 \\
				0 & b & (b-a^2) \\
				0 & 0 & a^2 
				\end{array}\right]}_{\boldsymbol{{A}}} \boldsymbol{\psi}(\boldsymbol{x})	.
			\end{equation}
			where not all coordinates are eigenfunctions i.e. ${x}_2$.
			The result of (\ref{OBmap}) can be then be projected onto the respective output observables of interest (system states) output matrix $\boldsymbol{C} = [\boldsymbol{I}_{2} \quad\boldsymbol{ 0}]$:
			\begin{equation}\label{eq:koopOF}
			[\boldsymbol{\mathcal{K}}_{\boldsymbol{F}}](\boldsymbol{x}) = \boldsymbol{F}(\boldsymbol{x}) = {\boldsymbol{C}}\boldsymbol{A}\boldsymbol{\psi}(\boldsymbol{ x}) .
			\end{equation}
			Here, the columns of the output matrix $\boldsymbol{C}$ (in system-theoretic sense) do not represent Koopman modes but by diagonalizing $\boldsymbol{K}$ into $\boldsymbol{\Lambda}$ (\ref{EFmap}), one uncovers Koopman operator eigenvalues, eigenfunctions and modes. As shown previously, one can opt to use $\boldsymbol{F}^{(k)}(\boldsymbol{x}) = {\boldsymbol{C}}\boldsymbol{A}^{k}\boldsymbol{\psi}(\boldsymbol{ x})$ for linear prediction as the formulations (\ref{eq:koopEF}) and (\ref{eq:koopOF}) indeed are equivalent.
			The notions from this example have their continuous-time analogue with the difference that one is capturing the effect of the Koopman generator operator instead.
		\end{enumerate}
	\end{example}	
}
\begin{figure}[ht!]
	\centering
	\makebox[0.7\textwidth]{\includegraphics[width=0.8\textwidth]{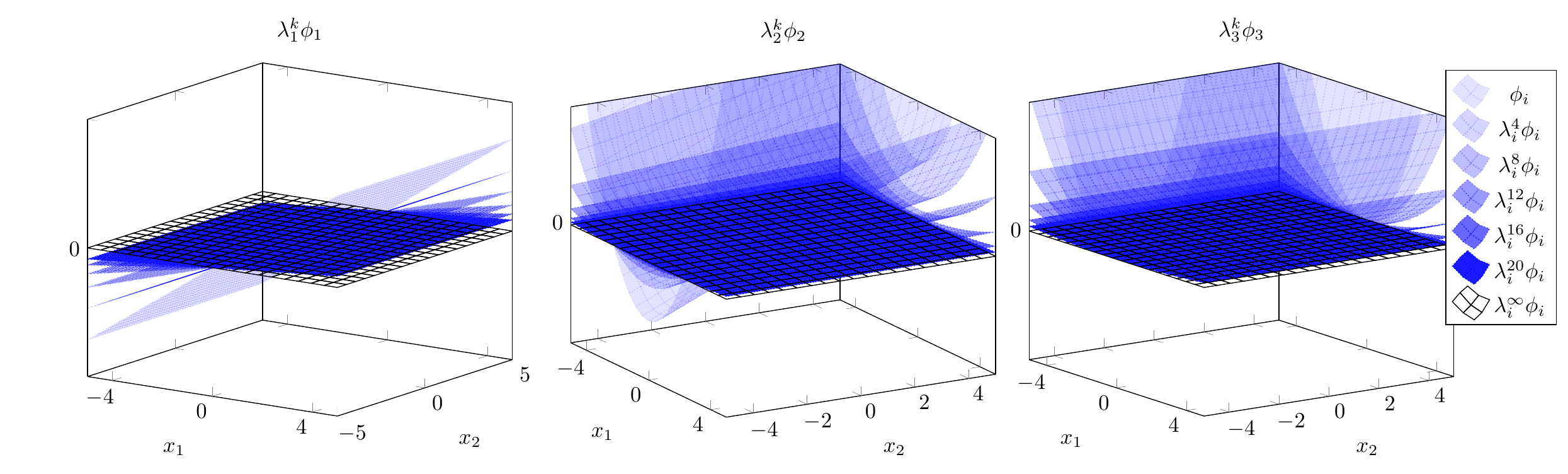}}
	\caption{Evolution of eigenfunctions over the state space from Example \ref{ex:motiv}\ref{ex:motivEF} for parameter values $a=0.9$ and $b=0.8$}
	\label{EFsliceEVO}
\end{figure} 
	{To illustrate the concept of linear systems representing nonlinear evolution in original coordinates we show the evolution of the respective eigenfunctions in Figure \ref{EFsliceEVO}. The linear combination of the linearly evolving eigenfunctions fully describes \textit{all} trajectories of the nonlinear system from Example \ref{ex:motiv}. This highlights the globality of the Koopman operator description of nonlinear systems and contrasts it to the local point dynamics study in originally nonlinear evolving coordinates of the $x_1$-$x_2$ plane. Already mentioned, the evolution of points in the state space is replaced by the evolution of hypersurfaces over the state space. Therefore, as all initial points start on the same initial surface (eigenfunctions) - they propagate linearly in a closed form.
\begin{figure}[ht!]
	\centering
	\makebox[0.5\textwidth]{\includegraphics[width=0.7\textwidth]{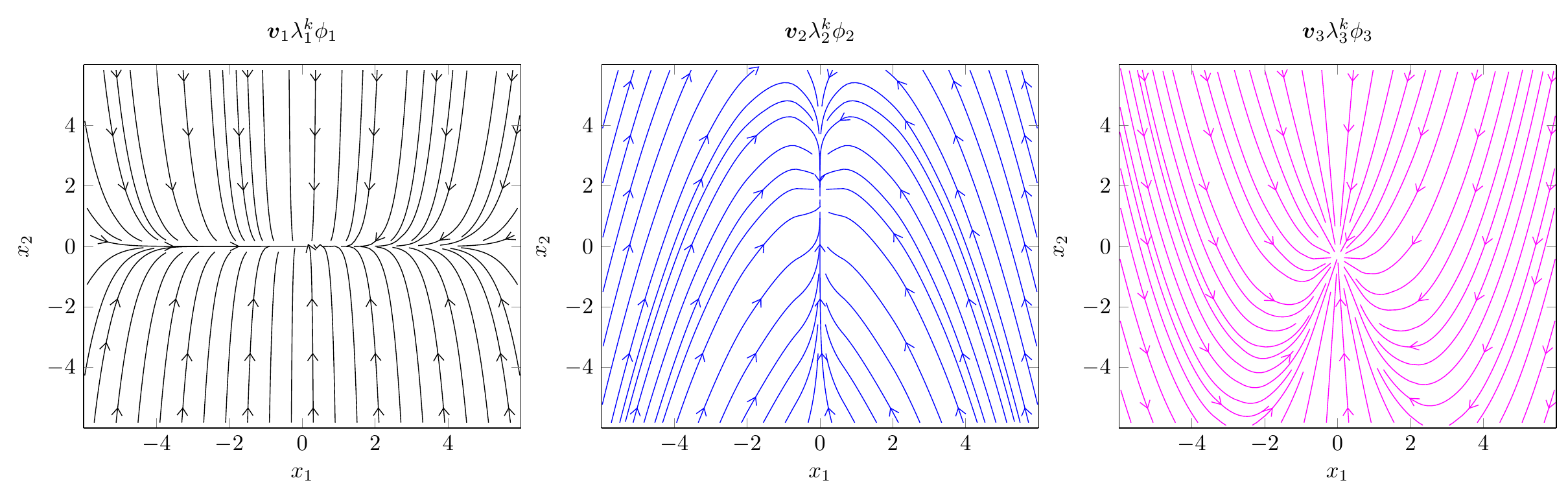}}
	\makebox[0.5\textwidth]{\includegraphics[width=0.5\textwidth]{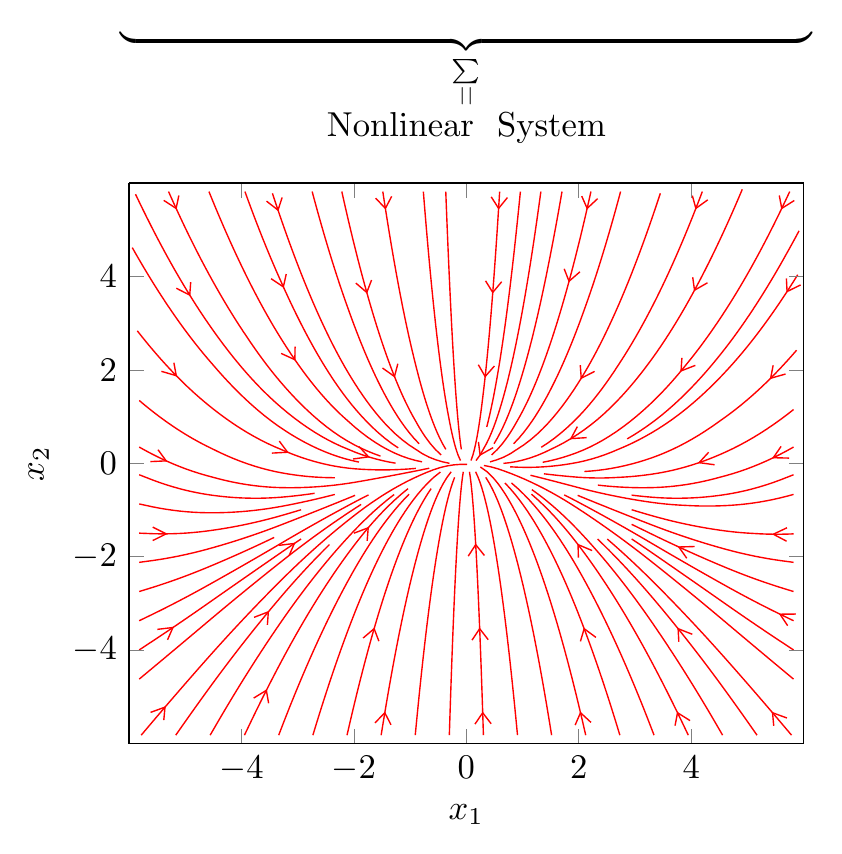}}
	\caption{The dynamical flows of the linearly evolving dynamical factors/modes superposed to form the nonlinear evolution in original coordinates $\boldsymbol{F}^{(k)}(\boldsymbol{x}) = {\boldsymbol{v}_1}{\lambda}_1^{k}{\phi}_1(\boldsymbol{x})+{\boldsymbol{v}_2}{\lambda}_2^{k}{\phi}_2(\boldsymbol{x})+{\boldsymbol{v}_3}{\lambda}_3^{k}{\phi}_3(\boldsymbol{x})$.}
	\label{ogNLSysEvo}
\end{figure}
In Figure \ref{ogNLSysEvo} one can recognize how the modal decomposition exposes the dynamical factors of the system that all evolve linearly and superposed give the original system's evolution. 
}

\subsection{Koopman operator's adjoint}
Finally, we remark how the Koopman operator has its adjoint (in appropriate function  spaces) - the Perron-Frobenius operator. As opposed to the Koopman operator that advances trajectories of a dynamical system forward in time, its Perron-Frobenius adjoint puts the dynamical effects into the density, pushing forward (probability) \emph{densities} over the state space \cite{ChaosBook} making it well suited for studying chaotic (stochastic) dynamical systems \cite[Chapter 4]{Lasota1994}.
\begin{property}[Koopman operator and its adjoint]\label{dualKP}
	{Given a measure $\mu$ on $\mathcal{M}$ and scalar-valued density $\rho \in L^{1}(\mathcal{M})$ observable ${\psi} \in L^{\infty}(\mathcal{M})$, the following holds
		\begin{equation}
		\left\langle \mathcal{P}^{t} \rho, \rho\right\rangle=\left\langle {\psi}, \mathcal{K}^{t} {\psi}\right\rangle,
		\end{equation}
		\begin{equation}
		\int_{\boldsymbol{x} \in \mathcal{M}}\left(\mathcal{ K }^{t} \rho(\boldsymbol{x})\right) \mu(d \boldsymbol{x})=\int_{\boldsymbol{x} \in \mathcal{M}} {\psi}(\boldsymbol{x}) \mathcal{P}^{t} \mu(d\boldsymbol{ x}) ,
		\end{equation}
		with $t \in \mathbb{R}_{+,0}$ where $\mathcal{P}^{t}$ denotes the semigroup of Perron-Frobenius (dual to Koopman) operators. } 
\end{property} 

Interestingly, Koopman and Perron-Frobenius operators are also referred to as \emph{backward} and \emph{forward} as they are solution operators of the backward and forward Kolmogorov (Fokker-Planck) equations, respectively \cite[Section 11]{Lasota1994}. Furthermore, the Perron-Frobenius operator generator $\mathcal{G}_{\mathcal{P}}$ of the stochastic differential equation $\dot{\rho}=\mathcal{G}_{\mathcal{P}}\rho$ corresponds to the Liouville (Fokker-Planck) operator.

\section{Data-driven Koopman operator-based dynamical models}\label{secIII}

{As the Koopman operator acts on a function space, it is infinite-dimensional in general. For a finite-dimensional nonlinear system, infinitely many dimensions might be needed to render it linear.  Thus, finding a suitable finite-dimensional representation of the operator is required. Predetermining a function basis on which the operator is projected represents the most widely used approach for finite-dimensional realizations. However, for an arbitrary basis type one might need infinitely many basis functions to have the operator closed in that space. By choosing a basis that is not arbitrary - but one that is closed under the operator - infinite dimensions are not required for a Koopman-invariant basis. Nevertheless, any set of Koopman-invariant coordinates is not guaranteed to reconstruct the states of interest well. Even with Koopman-invariant coordinates, one might need an infinite amount for exact state reconstruction - requiring truncation. Therefore, learning ``good'' coordinates (\emph{lifting} functions) that are genuinely linear and are able to reconstruct the states of interest is non-trivial.
The aforementioned reality lead to many different learning approaches for representing the Koopman operator. In the following we review the data-driven approaches for learning \textit{Koopman operator dynamical models} found in the literature.
}

\subsection{Origins and classification of Koopman operator discovery methods}\label{introDMD}
A large class of algorithms for data-driven Koopman operator approximations is in some way related to dynamic mode decomposition (DMD) \cite{schmid_2010}. DMD had major success in fluid flow modeling, reduced order modeling (ROM) and Koopman operator approximations \cite{Tu2014}. 
For a comprehensive dimension reduction and dynamical systems view on DMD we point to \cite{doi:10.1137/1.9781611974508:DMDkutz}. We base our outline in this section on it, as it represents a linear approximation of the Koopman operator with respect to the system state and  many of the more advanced methods from the literature stemmed from it.
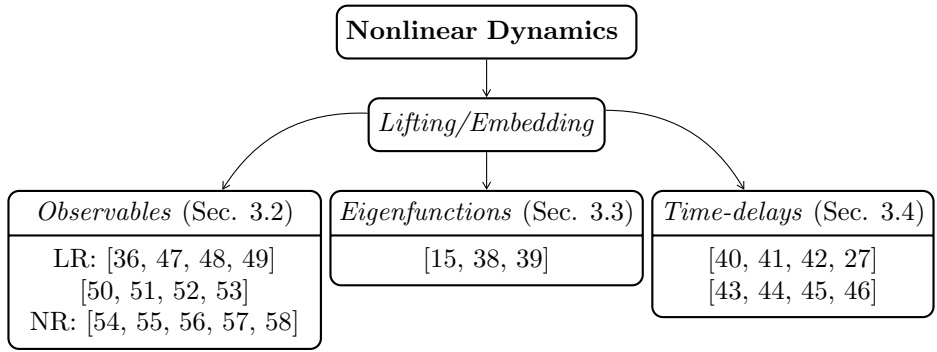
\begin{figure*}
	\centering
	
	\begin{tikzpicture}[
	node distance = 5mm and 6mm, 
	box/.style = {draw, rounded corners, 
		minimum width=22mm, minimum height=7mm, align=center,draw=black, thick,},
	> = {Straight Barb[angle=60:3pt ]},
	bend angle = 30,
	auto = right,
	state/.style={
		rectangle split,
		rectangle split parts=2,
		rectangle split part fill={red!0,blue!0},
		rounded corners,
		draw=black, thick,
		minimum height=30mm,
		minimum width=22mm,
		inner sep=3pt,
		text centered,
	}
	]
	\node (n1)  [box] { \textbf{Nonlinear} \textbf{Dynamics}  };
	\node (n3)  [box, below=of n1]  
	{\textit{Lifting/Embedding}};

	\node (n7)  [state, below =of n3,fill=blue!0]    
	{\textit{Eigenfunctions} (Sec. \ref{eigLift}) \nodepart{two}   $\begin{array}{c}
		\text{\cite{Korda2020b,Folkestad2019,Kaiser2017a}}
		\end{array}$ };
	\node (n5)  [state, below right =of n3,fill=red!0] 
	{\textit{Time-delays} (Sec. \ref{tdLift}) \nodepart{two}   $\begin{array}{c}
		\text{\cite{{Susuki2015},{LeClainche2017},Brunton2017a,Arbabi2017}}\\\text{\cite{Giannakis2019b,Kamb2018,Das2019a,Pan2019}}
		\end{array}$ };
	\node (n6)  [state, below left =of n3,fill=blue!0]   
{\textit{Observables} (Sec. \ref{obsLift}) \nodepart{two}  $\begin{array}{c}
	\text{LR: \cite{schmid_2010, Williams2015a,Williams2015b,Li2017a}}\\ \text{\cite{Huang2018, Yeung2019,Lian2019, Klus2020}} \\ \text{NR: \cite{Takeishi2017a,Lusch2018,Otto2019,Morton2019,Pan2020}}\end{array}$ };

	\draw[->] (n1) to [ ]  (n3);

	\draw[->] (n3) to [bend left,""]  (n5);
	\draw[->] (n3) to [bend right,""]  (n6);
	\draw[->] (n3) to [""]  (n7);
	\end{tikzpicture}
	\caption{A conceptual graph of the main branches of model representations employing the Koopman operator paradigm: \textit{root} - original system, \textit{level 1} - transformation type, \textit{level 2} - representation coordinates employed. The bottom split part of the latter includes relevant references employing the methodology noted in the top part. LR and NR denote linear/nonlinear reconstruction of the state, respectively.}
	\label{fig:graph}
\end{figure*}
The key ideas of DMD methods are closely related to model-reduction techniques and showed promise in data-based modeling of fluid flows \cite{Rowley2009a}. 
Briefly, {DMD} aims to extract dominant spatio-temporal modes from data, providing data-driven local-linear analysis of nonlinear systems which made it one of the first Koopman operator approximation methods.
Consider the data being formatted into `{snapshots}': 
\begin{align}
\boldsymbol{H}^{1}(\boldsymbol{x})&=\left[{\boldsymbol{x}_{0}},  {\cdots}, {\boldsymbol{x}_{T-1}} \right], \label{dmd1} \\
{ }^{+}\boldsymbol{H}^{1}(\boldsymbol{x})&=\left[ {\boldsymbol{x}_{1}}, {\cdots}, {\boldsymbol{x}_{T}} \right], \label{dmd1+}
\end{align} that are vector Hankel matrices with depth one of a $T$-step time-sequence.
Then the best-fit matrix $\boldsymbol{A}$ is obtained by minimizing 
\begin{equation}\label{dmdEQ}
\min_{\boldsymbol{A}}\left\|{}^{+}\boldsymbol{H}^{1}(\boldsymbol{x})-\boldsymbol{A}\left[\boldsymbol{H}^{1}(\boldsymbol{x})\right]\right\|_{F}
\end{equation} with index $F$ denoting the Frobenius norm. In closed form, the transition matrix is obtained as $\boldsymbol{A}={}^{+}\boldsymbol{H}^{1}(\boldsymbol{x})[\boldsymbol{H}^{1}(\boldsymbol{x})]^{\dagger}$ -
where ${\dagger}$ represents the Moore-Penrose pseudo-inverse. The utility of DMD for reduced order modeling for high-dimensional snapshots involves taking the singular value decomposition of the snapshot matrix $\boldsymbol{H}^{1}(\boldsymbol{x})$.

Regarding linear representations of nonlinear systems we focus on methods based on predicting \emph{output functions} of interest constructing a surrogate model of a system based on the Koopman operator paradigm. 
{
As depicted in Figure \ref{fig:graph}, we classify the methods employed to predict the system's outputs of interest into three categories based on the lifting coordinates employed. Note that this is only a tentative classification for better exposition as the methodologies are more closely related than Figure \ref{fig:graph} would suggest. The firstly developed train of thought considers a span of {observables as coordinates} looking to approximate the infinite-dimensional Koopman operator - a \textit{spectrally indirect} family of approaches. Other methods seeking to find inherently Koopman-invariant coordinates consider direct discovery of {eigenfunction coordinates} - \textit{spectrally direct}. Worth noting is the fact that the latter methodologies deals with the point spectrum (recall Remark \ref{rmk:fullSPECT}). Considering on-attractor (post-transient) behavior, \emph{time-delay coordinates} of output functions (instead of only spatial coordinates) give a useful linear representation of the system at hand. All of these frameworks require sequential data e.g. from a system's trajectory (to capture the spectrum), as opposed to learning with non-sequential input-output samples encountered in function approximation.}

\subsection{Lifting to observables}\label{obsLift}
Throughout the years many extensions of DMD came along, notably extended DMD (EDMD) \cite{Williams2015a}. It involves - often heuristically predetermined - nonlinear maps of the data (e.g. radial basis functions or monomials), hoping to capture the nonlinearity in the dynamics. For that, the data-snapshots (\ref{dmd1}-\ref{dmd1+}) are \emph{lifted} through a feature map $\boldsymbol{{\psi}}(\cdot)$ resulting in
\begin{align}
\boldsymbol{H}^{1}_{\boldsymbol{\psi}}(\boldsymbol{x})&=\left[\boldsymbol{{\psi}}({\boldsymbol{x}_{0}}),  {\cdots}, \boldsymbol{{\psi}}({\boldsymbol{x}_{T-1}}) \right], \\ {}^{+}\boldsymbol{H}^{1}_{\boldsymbol{\psi}}(\boldsymbol{x})&=\left[ \boldsymbol{{\psi}}({\boldsymbol{x}_{1}}), {\cdots}, \boldsymbol{{\psi}}({\boldsymbol{x}_{T}}) \right],
\end{align} and compatible closed-form regression as in (\ref{dmdEQ}). In what follows, we outline a more general formulation for obtaining Koopman operator dynamical models motivated by the aforementioned idea.

For a data-set $\mathbb{D}=\{\boldsymbol{x}_i,\boldsymbol{y}_i\}_{i=0}^{T-1}$ with dynamics $\boldsymbol{y}_i=\boldsymbol{F}(\boldsymbol{x}_i)$, the goal is to solve the following optimization problem:

\begin{align}
\operatornamewithlimits{min}_{\boldsymbol{A}, \boldsymbol{C}, \boldsymbol{\psi}(\cdot)}  \sum^{T-1}_{i=0}&\overbrace{\left\|\boldsymbol{z}_{i+1}-\boldsymbol{A} \boldsymbol{z}_i\right\|^{2}}^{linearity}+\overbrace{\left\|\boldsymbol{y}_{i}-\boldsymbol{C}\boldsymbol{z}_{i}\right\|^{2}}^{reconstruction} \label{obj_opt} \\
\text{subject to:} &  \quad \boldsymbol{z}_i=\boldsymbol{\psi}(\boldsymbol{x}_i) \label{lift_opt} \\ & \quad \boldsymbol{\psi} \in \mathcal{F} \label{space_opt}
\end{align}
where $\mathcal{F}$ is a space of observables.
Notice that this becomes equivalent to vanilla DMD (\ref{dmdEQ}) where there is no ``lifting" - $\boldsymbol{\psi}(\boldsymbol{ x})=\boldsymbol{x}$.
The targeted representation is in the form of a \emph{Koopman operator dynamical model}
\begin{align}
\boldsymbol{z}_{0} &=\boldsymbol{\psi}(\boldsymbol{x}_0), \label{LTIo:1}\\
\boldsymbol{z}_{k+1} &=\boldsymbol{A} \boldsymbol{z}_{k}, \label{LTIo:2}\\
\boldsymbol{y}_{k} &=\boldsymbol{C} \boldsymbol{z}_{k}, \label{LTIo:3}
\end{align}
where the initial lift (\ref{LTIo:1}) leads to linear state-transition coordinates (\ref{LTIo:2}) which are then projected on outputs of interest (\ref{LTIo:3}). 
Some EDMD variations and extensions include kernel EDMD \cite{Williams2015b} (using kernels instead of fixed basis functions) with its generator version - gEDMD \cite{Klus2020} and naturally structured EDMD \cite{Huang2018} (preserving the Markov property leading to stable finite-dimensional approximations). 
However, the aforementioned EDMD approaches fix a lifting map (\ref{lift_opt}) and then solve (\ref{obj_opt}) as an convex, least-squares optimization problem. The unstructured fixed-bases of the above EDMD methods often lead to data-inefficient models of very high dimension and only locally accurate prediction. To tackle the aforementioned issue, Li et al. \cite{Li2017a} propose simultaneous learning of (\ref{lift_opt}) with a neural network and an added $L_1$-regularizer in (\ref{obj_opt}). Furthermore, Lian and Jones \cite{Lian2019} propose the use of subspace identification methods \cite{Gustafsson2002} for solving (\ref{obj_opt}) independently of the lifting in (\ref{lift_opt}) which is subsequently fitted.

Note that all of the approaches mentioned until now are only able to represent purely transient (off-attractor) dynamics, as there exists a finite-dimensional Koopman-invariant subspace that includes the state itself for (almost) global linear representations.
To be able to represent a wider range of dynamical regimes, some approaches are considering a nonlinear reconstruction map to the original state instead of a linear one (\ref{LTIo:3}) requiring invertibility and a suitably modified reconstruction loss compared to (\ref{obj_opt}). Such a modification does not require the state to lie in the \textit{span} of lifted coordinates and can lead to lower-dimensional embeddings as well. To discover such coordinates, machine learning tools such as neural network structures in various forms, e.g. linear-recurrence and auto-encoding \cite{Takeishi2017a,Otto2019,Morton2019,Pan2020} are employed.
To tackle continuous spectra as well, the work of Lusch et al. \cite{Lusch2018} frequency-parameterizes a low-dimensional Koopman embedding using neural networks. Although nonlinear reconstructions can lead to lower-dimensional representations, siding with \emph{linear reconstruction} is useful for practical control design and amendable linear mode analysis of nonlinear systems. 

{Considering a collection of observables (not necessarily eigenfunctions) as lifting coordinates is still the most dominant train of thought in the data-driven frameworks for Koopman operator-based representations. Nonetheless, such approaches often lead to lifting coordinates not being in the span of eigenfunctions - providing only locally accurate prediction. Furthermore, unstructured operator approximation often does not deliver approximation or convergence to the real eigenvalues/eigenfunctions as the approaches are \textit{spectrally indirect}. The aforementioned introduces dissonance into the analysis procedures for such models because the spectrum is not close to the real one and thus is misrepresenting the properties of the underlying system. Recalling Example \ref{ex:motiv}, one can recognize that the problem of directly looking for eigenfunctions (\textit{spectrally direct}) is better posed for finding genuine linearly-evolving coordinates as there is more structure that can be advantageously exploited. Thus, if one looks to exploit the paradigm for long-term prediction as well as analysis, considering learning the Koopman operators in a spectrally direct fashion can be more efficacious.}
\subsection{Lifting to eigenfunction coordinates}\label{eigLift}
In the following we present Koopman-operator representation approaches that are not purely data-based but also \emph{structurally aware} - looking to exploit the Koopman operator's algebraic and geometric properties.
Due to a strong connection of Koopman operators and state space geometry, exploiting notions from differential geometry can provide, to a degree, ``supervision" to the generally unsupervised task of learning Koopman operator eigenfunctions.

\subsubsection{Theoretical basis}
As opposed to other approaches that purely use data to obtain a span of - often not truly Koopman-invariant - observables, the focus here is on using (generalized) eigenfunctions \emph{by design} relevant for predicting {output functions} which calls for a more rigorous theoretical treatment.
\begin{theorem}[\cite{Bollt2017}]\label{charKG}
	Consider the domain $\mathbb{X} \subseteq \mathcal{M} \subseteq \mathbb{R}^n$, $\boldsymbol{x} \in \mathbb{X}$ and $\dot{\boldsymbol{x}}=\boldsymbol{f}(\boldsymbol{x})$ such that $\boldsymbol{f}: \mathbb{X} \mapsto \mathcal{M}$, then the
	corresponding Koopman operator has eigenfunctions $\phi(\boldsymbol{x})$ that are solutions of a linear partial differential equation (PDE)
	$$
		\frac{\partial \phi}{\partial \boldsymbol{x}} \boldsymbol{f}(\boldsymbol{ x})=s \phi(\boldsymbol{x})
	$$
	if $\mathbb{X}$ is compact and $\phi(\boldsymbol{x}) \in C^1(\mathbb{X})$ is in $C^{1}(\mathbb{X})$, or alternatively, if $\phi(\boldsymbol{x})$ is $C^{2}(\mathbb{X})$.
\end{theorem}

To highlight the connection of Koopman eigenfunctions with algebraic and geometric notions, we base our exposition on Definition \ref{linCoor} that, when fixing time-$t$, turns into an eigenvalue problem 
\begin{equation}
\phi\left( \boldsymbol{F} (\boldsymbol{x}) \right) = \lambda \phi \left(\boldsymbol{x} \right).
\end{equation}
Interestingly, the univariate version of the above equation is the Schr\"oder's functional equation $\phi \circ F = \lambda \phi$ \cite{Schroder1870}. This equation studies how maps evolve under iteration and even pre-dates Koopman's seminal work \cite{Koopman1931}. 
Considering the Schr\"oder's equation in its multivariate form \cite{Cowen2003}, we write
\begin{equation}\label{Schroeder}
\boldsymbol{\mathcal{K}}_{\boldsymbol{F}} \boldsymbol{\phi} = \boldsymbol{\phi} \circ \boldsymbol{F} = \boldsymbol{F}^{\prime}(\boldsymbol{0}) \boldsymbol{\phi},
\end{equation}
with $\boldsymbol{F}^{\prime}(\boldsymbol{0})$ being the linearization of $\boldsymbol{F}: \mathbb{X} \mapsto \mathcal{M}$ around the origin.
The full rank (univalent) solution of (\ref{Schroeder}) is analytic if $0<||\boldsymbol{F}^{\prime}(\boldsymbol{0})||<1$ \cite{Bridges2011} and delivers nontrivial \emph{principle} eigenfunctions (Definition \ref{princEP}) of the map $\boldsymbol{F}$. Schr\"oder's equation (\ref{Schroeder}) is also directly related to {functional conjugation} \cite{Curtright2011} on whose implications we elaborate on in the following. 

\begin{figure}[ht!]
	\centering
	
	\begin{tikzpicture}[baseline= (a).base]
	\node[scale=1.2] (a) at (0,0){
		\begin{tikzcd}[column sep = huge, row sep = huge]
		\mathcal{M} \arrow{r}{\boldsymbol{F}} \arrow[swap]{d}{\boldsymbol{d}} & \mathcal{M} \arrow{d}{\boldsymbol{d}} \\%
		\mathcal{Y} \arrow{r}{\boldsymbol{T}}& \mathcal{Y}
		\end{tikzcd}
	};
	\end{tikzpicture}

	\caption{Commutative diagram of topological conjugacy} 
	\label{fig:conj}
\end{figure}
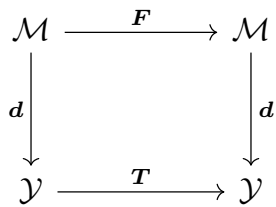

\begin{definition}[Topological conjugacy]
	Consider maps $\boldsymbol{F}: \mathcal{M} \mapsto \mathcal{M}$ and $\boldsymbol{T}: \mathcal{Y} \mapsto \mathcal{Y}$ such that $\exists {\boldsymbol{d}}: \mathcal{M} \mapsto \mathcal{Y}$ satisfying $\boldsymbol{d} \circ \boldsymbol{F} = \boldsymbol{T} \circ \boldsymbol{d} $. Then, the maps are \emph{topologically} 
	\begin{enumerate}[label=\roman*)]
		\item \emph{conjugate} if $\boldsymbol{d}$ is {a homeomorphism}
		\item \emph{semi-conjugate} if $\boldsymbol{d}$ is {$C^0$ and surjective}
	\end{enumerate}
\end{definition}
\begin{remark}
	Consider the maps from Figure \ref{fig:conj}. If conjugated, then $\boldsymbol{F}$ and $\boldsymbol{T}$ have the same dynamic behavior while, if they are semi-conjugated, the dynamics of $\boldsymbol{T}$ are contained in $\boldsymbol{F}$. 
\end{remark}

\begin{theorem}[Spectral equivalence \cite{Budisic2012a}]\label{specEQ}
	Let the maps $\boldsymbol{F}: \mathcal{M} \mapsto \mathcal{M}$ and $\boldsymbol{T}: \mathcal{Y} \mapsto \mathcal{Y}$ be topologically conjugate via the homeomorphism ${\boldsymbol{d}}: \mathcal{M} \mapsto \mathcal{Y}$ such that $\boldsymbol{d} \circ	\boldsymbol{F} = \boldsymbol{T} \circ \boldsymbol{d} $ holds. If $(\lambda, \varphi)$ is an eigenpair of $\mathcal{K}_{\boldsymbol{T}}$, then $(\lambda, \varphi \circ \boldsymbol{d})$ is an eigenpair of $\mathcal{K}_{\boldsymbol{F}}$.
\end{theorem}

Lan and Mezi\'{c} \cite{Lan2013} consider extending the validity of the conjugacy via the Hartman-Grobman theorem to the whole basin of attraction. The work of Mohr et al. \cite{Mohr2016} shows that a nonlinear system's eigenfunctions can be constructed by Theorem \ref{specEQ} from eigenpairs of its asymptotically stable liearization. For a more detailed continuous-time treatment on conjugacy with connections to matching and rectification of vector fields we point to \cite{Bollt2017}. 
\begin{example}\label{ex2}
	Consider again the system from Example \ref{ex:motiv}
	\begin{equation}
	\boldsymbol{x} \mapsto \boldsymbol{F}(\boldsymbol{x}) = \left[\begin{array}{c}
	a x_{1} \\
	b x_{2}+\left(b-a^{2}\right) x_{1}^{2}
	\end{array}\right],
	\end{equation}
	and its linearization
	\begin{equation}
	\boldsymbol{y} \mapsto \underbrace{\left[\begin{array}{cc}
		a & 0  \\
		0 & b  
		\end{array}\right]}_{\boldsymbol{T} = \boldsymbol{F}^{\prime}(\boldsymbol{0})}\boldsymbol{y}.
	\end{equation}
	Evidently, the principle eigenpairs of the linearized system are $(a,\left\langle \boldsymbol{w}_a, \boldsymbol{y} \right\rangle)$ and $(b,\left\langle \boldsymbol{w}_b, \boldsymbol{y} \right\rangle)$ with $\boldsymbol{w}_{a}$, $\boldsymbol{w}_{b}$ the eigenvectors of the adjoint of $\boldsymbol{T}$ \cite{Mohr2014}. By Theorem \ref{specEQ} we see there is a conjugacy map $\boldsymbol{d}(\boldsymbol{x})=[x_1,x_2+x^2_1]^{\top}=\boldsymbol{x}+\boldsymbol{h}(\boldsymbol{ x})$ (Figure \ref{fig:conj}) with the diffeomorphism's residual $\boldsymbol{h}(\boldsymbol{ x})=[0,x^2_1]^{\top}$ allowing for construction of non-trivial principal eigenfuncions of the nonlinear system from the ones of its conjugate linear system. {Recognize how the map $-\boldsymbol{h}(\boldsymbol{x})$ defines the invariant manifolds of the full nonlinear system (see Figure \ref{invManPic}).
\begin{figure}
	\centering
	\makebox[0.75\textwidth]{\includegraphics[width=0.75\textwidth]{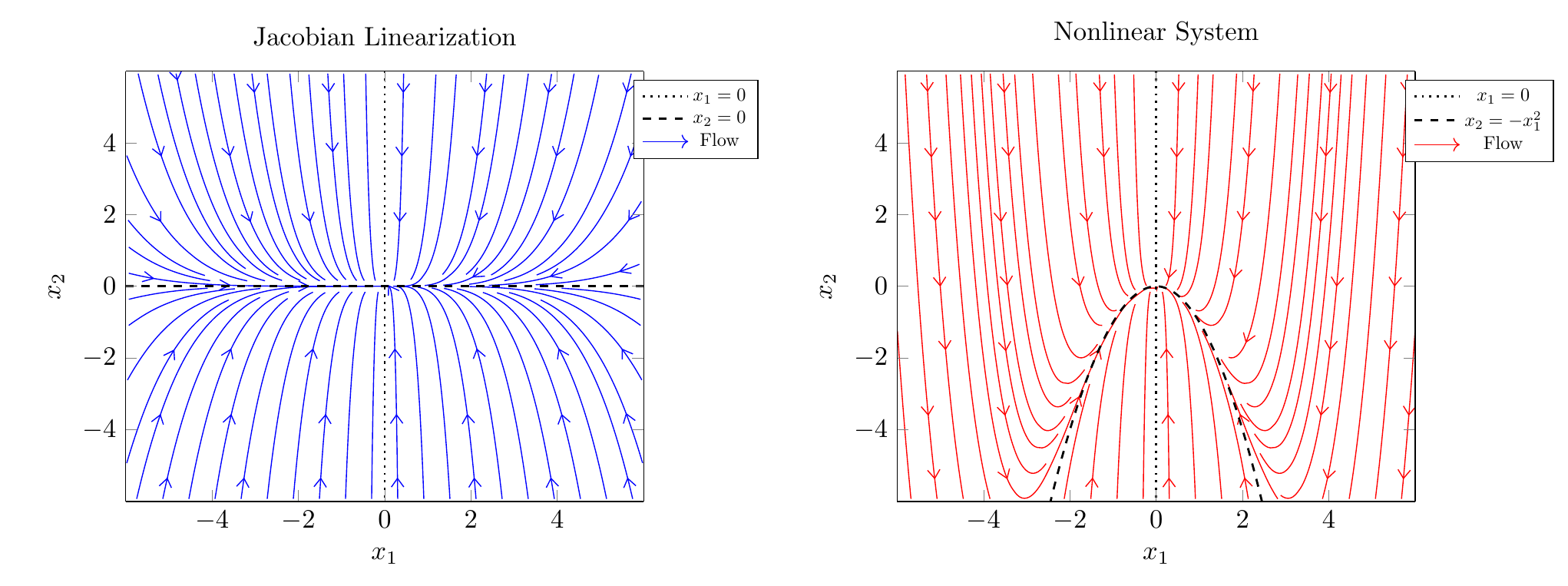}}
	\caption{Warping the linear system to a nonlinear one through topological conjugacy for Example \ref{ex:motiv} with parameters $a=0.9$ and $b=0.5$.}
	\label{invManPic} 
\end{figure}
}	
\end{example}

\begin{remark}[Semi-conjugacy and ROM]
	Although for general observables infinite-dimensional, the composition operator is one-dimensional for its eigenfunctions which are semi-conjugacies by definition $\lambda \circ {\phi} ={\phi}\circ{\boldsymbol{F}}$. This realization theoretically justifies the utility of the Koopman operator paradigm in reduced order modeling (ROM) such as \cite{Arbabi2017,Peitz2019,Klus2018a,Klus2020}.
\end{remark}

\subsubsection{Data-driven approaches}
The data-driven frameworks utilizing the outlined theory assume the following Koopman operator dynamical model
\begin{align}
\boldsymbol{z}_{0} &=\boldsymbol{\phi}(\boldsymbol{x}_0), \label{LTI:1}\\
\boldsymbol{z}_{k+1} &=\boldsymbol{\Lambda} \boldsymbol{z}_{k}, \label{LTI:2}\\
\boldsymbol{y}_{k} &=\boldsymbol{V} \boldsymbol{z}_{k}. \label{LTI:3}
\end{align}
where $\boldsymbol{\phi}$ is (generalized) eigenfunction lifting with $\boldsymbol{\Lambda}$ being the eigenfunction transition map and $\boldsymbol{V}$ the projection on the outputs (Koopman modes).

Utilizing the introduced structural properties leads to a modified optimization objective compared to (\ref{obj_opt})-(\ref{space_opt}). Formally, we can write
\begin{align}
\operatornamewithlimits{min}_{\boldsymbol{\Lambda}, \boldsymbol{V}, \boldsymbol{\phi}(\cdot)}  \sum^{T-1}_{i=0}&\left\|\boldsymbol{z}_{i+1}-\boldsymbol{\Lambda} \boldsymbol{z}_i\right\|^{2}+\left\|\boldsymbol{y}_{i}-\boldsymbol{V}\boldsymbol{z}_{i}\right\|^{2} \label{obj_optEF} \\
\text{subject to:} &  \quad \boldsymbol{z}_i =\boldsymbol{\phi}(\boldsymbol{x}_i) \label{lift_optEF} \\ 
& \quad \boldsymbol{\Lambda} \in \mathcal{S}_{\lambda}, \label{constr_optEV} \\
& \quad \boldsymbol{\phi} \in \mathcal{S}_{{\phi}}, \label{constr_optEF}
\end{align}
where $\mathcal{S}_{\phi}, \mathcal{S}_{\lambda}$ represent \emph{structural} constraints for eigenfunctions and eigenvalues, respectively. This leads to a better-posed problem with a more restricted solution space than in (\ref{space_opt}). In order to provide eigenfunction lifting by design some geometrical and topological considerations need to be made, which we present in the following. 
\paragraph{Learning via topological conjugacy}
{ The assumed system form consists of a known linear part and an unknown residual nonlinearity
\begin{equation}\label{sysKdiffeo}
\dot{\boldsymbol{x}}=\boldsymbol{f}(\boldsymbol{x})=\boldsymbol{T}\boldsymbol{x}+\boldsymbol{r}(\boldsymbol{x}).
\end{equation}
Utilizing topological conjugacy from Theorem \ref{specEQ} takes the role of constraints (\ref{constr_optEV})-(\ref{constr_optEF}) - constraining the choice of eigenvalues and coordinate-bases to genuine Koopman eigenfunctions.
The conjugacy arises as a solution of the following PDE
\begin{equation}\label{conjPDE}
\begin{gathered}
\boldsymbol{T} \boldsymbol{h}(\boldsymbol{x})-\boldsymbol{r}(\boldsymbol{x})=\frac{\partial}{\partial \boldsymbol{ x}}\boldsymbol{h}(\boldsymbol{x})(\boldsymbol{T} \boldsymbol{x}+\boldsymbol{r}(\boldsymbol{x})) \\
\boldsymbol{h}(\boldsymbol{0})=\boldsymbol{0}, \quad \frac{\partial \boldsymbol{h}}{\partial \boldsymbol{ x}}(\boldsymbol{0})=\boldsymbol{0}
\end{gathered}
\end{equation}
leading to the following equations via the method of characteristics
\begin{equation}
\label{charEQ}
\begin{aligned}
\dot{\boldsymbol{x}} &=\boldsymbol{T} \boldsymbol{x}+\boldsymbol{r}(\boldsymbol{x}) \\
\dot{\boldsymbol{z}} &=\boldsymbol{T} \boldsymbol{z}-\boldsymbol{r}(\boldsymbol{x})
\end{aligned}
\end{equation}
after defining $\boldsymbol{d}(\boldsymbol{x})=\boldsymbol{x} + \boldsymbol{h}(\boldsymbol{x})$ and $\boldsymbol{z}=\boldsymbol{h}(\boldsymbol{x})$ \cite{Wang2020b}.

\begin{remark}\label{origCONJ}
	Notice that considering the Koopman operator analogue of the infinitesimal (generator) solution (\ref{charEQ}) for the system form $\mathcal{K}_{\boldsymbol{F}}[\boldsymbol{x}] = \boldsymbol{A}\boldsymbol{x}+\boldsymbol{e}(\boldsymbol{x})$
	\begin{equation}
	\label{charEQdt}
	\begin{aligned}
	\mathcal{K}_{\boldsymbol{F}}[{\boldsymbol{x}}] &=\boldsymbol{A}\boldsymbol{x}+\boldsymbol{e}(\boldsymbol{x}) \\
	\mathcal{K}_{\boldsymbol{F}}[{\boldsymbol{z}}] &=\boldsymbol{A} \boldsymbol{z}-\boldsymbol{e}(\boldsymbol{x})
	\end{aligned}
	\end{equation}
	where recognizing $\mathcal{K}_{\boldsymbol{F}}[{\boldsymbol{z}}]= \boldsymbol{A}\boldsymbol{d}(\boldsymbol{x}) - \boldsymbol{F}(\boldsymbol{x})$ the principle eigenfunctions are dimension-wise components of the diffeomorphism $\boldsymbol{d}(\boldsymbol{x})=\boldsymbol{x}+\boldsymbol{h}(\boldsymbol{x})$ from Example \ref{ex2}. 
\end{remark}

 Thus, one can approximately learn the solution of (\ref{conjPDE}) by solving the following optimization problem based on the dynamics residual $\boldsymbol{r}(\boldsymbol{x})$ samples
\begin{align}
\operatornamewithlimits{min}_{\boldsymbol{h}(\cdot)}  \sum^{T-1}_{i=0}&\|\dot{\boldsymbol{d}}_{i}-\boldsymbol{T}\boldsymbol{d}_i\|^{2}  \\
\text{subject to:} & \quad \boldsymbol{d}(\boldsymbol{x}_{i}) = \boldsymbol{x}_{i}+\boldsymbol{h}\left(\boldsymbol{x}_{i}\right) \\
& \quad \frac{\partial}{\partial \boldsymbol{ x}}\boldsymbol{h}(\boldsymbol{0})=\boldsymbol{0} 	,
\end{align}
as pursued in \cite{Folkestad2019} in a similar manner where the goal is to learn an LTI system that can well approximate the Koopman operator generator.
Subsequently, one is able to have a recursive construction of eigenfunction coordinates from the principal ones following Property \ref{productEFs} and Theorem \ref{specEQ}. Using these properties takes the role of (\ref{constr_optEF}) for learning Koopman-based dynamical models by lifting to eigenfunction coordinates by design.}

\paragraph{Learning by exploiting state space geometry}
The approach of Korda et al. \cite{Korda2020b} optimizes \emph{generalized} eigenfunctions, changing (\ref{LTI:1})-(\ref{LTI:3}) to admit a continuous-time representation for the state evolution matrix to a block diagonal Jordan matrix that is then discretized.
Known from finite-dimensional operator theory, generalized eigenfunctions do give rise to block diagonal Jordan blocks 
\begin{equation}\label{JB}
\boldsymbol{J}_{\lambda}=\left[\begin{array}{lll}
\lambda & 1 \\
& \ddots & 1 \\
& & \lambda
\end{array}\right],
\end{equation}
that decompose the space of observables into invariant subspaces of the Koopman operator semigroup spanned by \emph{generalized eigenfunctions} \cite{Korda2020b}.
The approach focuses on multi-step prediction error minimization while using non-recurrent sets to construct {generalized eigenfunctions}. 
More specifically, the eigenfunctions are constructed by exploiting the existence of boundary functions - values of eigenfunctions on non-recurrent sets of the state space - which serve as a structural constraint one eigenfunctions in (\ref{obj_opt})-(\ref{space_opt}). It also needs a suitably rich choice of initial conditions and requires system trajectories for learning.
Nevertheless, the method does not require any prior knowledge about the system and predicates solely on convex optimization methods. 

We also highlight some works not geared towards representing the original nonlinear system as (\ref{LTI:1})-(\ref{LTI:3}) but do seek to find solutions to the eigenvalue equations (\ref{KGEVE}) and (\ref{genEF}) for i.e. coherent structure extraction \cite{Klus2018,Klus2019} or discovering conservation laws \cite{Kaiser2019}.
When it comes to identifying Koopman operator principle eigenpairs, the work of Klus et al. \cite{Klus2020a} (generator version \cite{Klus2020}) employs eigendecompositions of Koopman operator using kernel methods.
On the Koopman operator generator identification front, the work of Kaiser et al. \cite{Kaiser2018} discovers only lightly-damped Koopman generator eigenpairs (as vanilla DMD-based methods converge to the strongest magnitude eigenvalues \cite[Section 2]{Budisic2012}). {However, the aforementioned approaches \cite{Klus2018,Klus2019,Kaiser2019,Klus2020a,Kaiser2018} highly depend on the empirical choice of basis function or kernel parameters.}

\subsection{Time-delay coordinates}\label{tdLift}
While the two previously introduced methodologies considered spatial coordinates, for ergodic systems, we turn to delay-coordinates approaches often employed for time-series prediction. These ideas are highly represented in the literature as the original Koopman theory \cite{Koopman1931} is used to prove Birkoff's ergodic theorem \cite{Koopman1932}.
In case that post-transient dynamics are of interest, purely spatial coordinates (Remark \ref{rmk:fullSPECT}) are of limited relevance on chaotic attractors. It is known from Takens embedding theory \cite{Takens1998} that one can provide a geometric reconstruction of a (strange) attractor of nonlinear systems by embedding scalar output measurement (observation) $y(k):=y_k$ into a higher dimensional space of time-delay coordinates.
This viewpoint is naturally defined for sampled-data systems.
	Here the observable (measurement) coordinates are based on the time-series $\left\{y_{k}\right\}_{k=0}^{T-1}, T \in \mathbb{N}$ of scalar measurements $y_k=h(\boldsymbol{x}_k)$ represented through a Hankel matrix
	\begin{equation}\label{HankM}
	\boldsymbol{H}^{L}(y):=\left[\begin{array}{cccc}
	y_{0} & y_{1} & \ldots & y_{T-L} \\
	y_{1} & y_{2} & \ldots & y_{T-L+1} \\
	\vdots & \vdots & \ddots & \vdots \\
	y_{L-1} & y_{L} & \ldots & y_{T-1}
	\end{array}\right],
	\end{equation}
	with \emph{depth} $L \in \mathbb{N}$. The argument of the Takens delay-embedding theorem is that an attractor of a dynamical system can be reconstructed using sufficiently time-delayed measurements ${\boldsymbol{y}}_{k}^{N}$ of a scalar observable of interest. The evolution of time-series sequences ${\boldsymbol{y}}_{k}^{N} \mapsto {\boldsymbol{y}}_{k+1}^{N}$ of the form 
	\begin{equation}
	{\boldsymbol{y}}_{k}^{N}:=\left[\begin{array}{cccc}
	{y}_{k} & \mathcal{K} {y}_{k} & \cdots & \mathcal{K}^{N} {y_{k}}\\
	\end{array}\right]^{\top} \in \mathbb{R}^{N \leq L},
	\end{equation} 
	becomes almost linear even when the measurement evolution $y_k \mapsto y_{k+1}$ comes from a nonlinear ergodic system \cite{Brunton2017a}.

Promising results of \cite{Kamb2018} use convolutions coordinates as a generalization of Hankel alternative view of the Koopman (HAVOK) framework of Brunton et al. \cite{Brunton2017a}. The latter work demonstrated the ability to construct a linear system with intermittent forcing that is able to predict chaotic dynamics.
Other notable extensions and generalizations of DMD with time-delay coordinates include Hankel-DMD \cite{Arbabi2017}, Prony approximation of Koopman operator \cite{Susuki2015} and higher order DMD \cite{LeClainche2017} that employ a time-delayed observable as a basis for DMD. Work of Arbabi et al. \cite{Arbabi2017} establishes convergence of Hankel-DMD to the true Koopman eigenfunctions and eigenvalues of the system. A notable use of combining both spatial observables and delay-embeddings can be found in Arbabi et al. \cite{Arbabi2019} for modeling nonlinear PDEs.
However it is known that linear models such as those generated by the Hankel-DMD can reconstruct an ergodic dynamical system only in an asymptotic sense. A study of this limitation and the conservatism of Takens theorem regarding the number of time-delays necessary can be found in \cite{Pan2019}.
As proven to useful for Hamiltonian systems and post-transient behaviors of system with quasi-periodic dynamics and continuous spectra (cf. Remark \ref{rmk:fullSPECT}), the utility of this method is in the literature have been reserved to ergodic (measure-preserving) systems. For a more detailed data-driven treatment of ergodic systems and Koopman spectra (beyond time-delay coordinates) we point to the \cite{Das2019a,Giannakis2019b}.

Time-delays as they pertain to the Hankel matrix (\ref{HankM}) also form the backbone of common linear-system identification such as the Ho-Kalman algorithm \cite{Ho1966} and the family of \emph{subspace state space identification} (4SID) methods \cite{Qin2007}. Although suggested to be similar to dynamic mode decomposition on a few occasions \cite{Tu2014, doi:10.1137/1.9781611974508:DMDkutz}, subspace-identification is fundamentally different from DMD.
We elaborate on this notion in the following remark. 
\begin{remark}[DMD vs 4SID]	
	In essence, ``vanilla''  DMD is a regression procedure linearly mapping two data `snapshots' (sequence of spatial measurements, Hankel matrices, images, etc.) from instance $k \mapsto k+1$ (allowing for dimension reduction via SVD \cite{doi:10.1137/1.9781611974508:DMDkutz}). On the other hand, 4SID is an unsupervised learning method for causal linear systems that explicitly delivers state space matrices (\ref{LTIo:2})-(\ref{LTIo:3}) via the system-theoretic concept of observability \cite{QIN20052043} while not providing the coordinate transform of (\ref{LTIo:1}). Although both methodologies might perform an SVD of the same Hankel matrix in the case of i.e. Hankel-DMD \cite{Arbabi2017} or HAVOK \cite{Brunton2017a}, they are fundamentally different and generally deliver different quantities. 
	Nevertheless, for a Hankel matrix \eqref{HankM} of depth one, 4SID methods do become equivalent to vanilla DMD and thus equally as ill-conditioned \cite[Section 2]{Budisic2012}.
\end{remark}
Another family of approaches considers approximating the Koopman operator generator through advection-diffusion operators \cite{Berry2016,Giannakis2019b} and time-delay coordinates. Such approaches relate to manifold learning and diffusion coordinates \cite{Coifman2006} as the data of the post-transient dynamics on an attractor forms a manifold. As already mentioned, approaches here recover Koopman operator-related quantities asymptotically. Also a recent approach of Takeishi et al. \cite{Takeishi2019} uses the fact that the diffusion operator commutes with the Koopman operator asymptotically and then constrains the learning algorithm to ensure best-possible commuting for the finite-delay and finite-data case.
\begin{remark}[Delay-coordinates and immersion]
	Recently, in \cite{Wang2020} the authors propose a relation of dynamical \emph{immersion} to time-delay coordinates in Koopman operator theory. There, the immersion is created by augmenting the full state with compositions of its state-transition map (delay-embedding). Such immersion-based coordinates are shown to be useful in the context of transient dynamics akin the frameworks in Section \ref{obsLift}. While conceptually similar, the immersion-based time-delays are of the full state and assume a known state transition map - different to the delay-embeddings considered in this section.
\end{remark}	
\subsection{Robotics applications}
With a focus on existing robotics applications, here we consider some notable data-driven approaches, exploring the practical utility of the Koopman operator paradigm.
A quite natural application can be found in soft robotics. The reason is obvious - with soft robots we do not even know what the states of interest are, let alone how to get them in a tractable way. That makes the Koopman operator dynamical models an interesting choice for data-driven control of soft robotic systems \cite{Bruder2019a,Bruder2020}.
Notably, learning fast quad-rotor landing of Folkestad et al. \cite{Folkestad2020} represents one of few data-driven methods with a strong
operator-theoretic motivation for Koopman operator dynamical models in robotics - proposing to learn a diffeomorphism to discover
eigenfunction coordinates for linear control design and prediction

\section{System analysis and Koopman operator paradigm}\label{secIV}

Representations of nonlinear systems via the ``dynamics of states" (\ref{minSys}) do not offer any intrinsic information relevant for system analysis. However, approaches using the Koopman paradigm, contain `free' information about the system of interest - invariant manifolds, fixed points, attractors and periodic orbits. The aforementioned quantities encode a great deal of information about the system and are the basis for system analysis through the Koopman operator paradigm.
\subsection{Classical stability analysis and Koopman paradigm}

In the following, we firstly highlight a natural connection of the Koopman paradigm to classical system-theoretic analysis through the generator of the Koopman operator semigroup.
In this classical view, we use the fact that the Lyapunov function itself is a special kind of an observable \cite{Mauroy2013,Mauroy2013,Mauroy2013b,Mauroy2020} with a negative semi-definite derivative. With that in mind, one can argue that the Koopman operator notions have implicitly been a part in Lyapunov stability analysis throughout the last century.
Consider the deterministic generator of the Koopman operator semigroup, already introduced in Definition \ref{generator}. 
{\begin{theorem}[\cite{khalil2002nonlinear}]\label{CTdetOP}
	Let there exist a proper Lyapunov function $V \in C^{2}$ associated with the system (\ref{minSys}) for which the action of the Lie-derivative operator $\mathcal{L}_{\boldsymbol{f}}$ along the vector field $\boldsymbol{f}$ results in
	\begin{equation}
	\mathcal{L}_{\boldsymbol{f}}V (\boldsymbol{ x})=\frac{\partial V}{\partial \boldsymbol{x}} \boldsymbol{f}(\boldsymbol{ x})  < 0,
	\end{equation}
	which holds for $\forall \boldsymbol{x} \in \mathcal{M} \backslash \{\boldsymbol{x}_*\}$. Then, the equilibrium point $\boldsymbol{x}_*$ of (\ref{minSys}) is asymptotically stable.
\end{theorem}
}
\begin{remark}\label{CTstab}
	A Lyapunov function can be seen as a non-negative \textit{observable} such that it is negative-semidefinite under the action of the Koopman operator generator $\mathcal{G}_{\mathcal{K}_{\boldsymbol{f}}}$ of the Koopman operator family $\{\mathcal{K}^{t}_{\boldsymbol{f}}\}_{t \in \mathbb{R}_{+,0}}$
	\begin{equation}
	\mathcal{G}_{\mathcal{K}_{\boldsymbol{f}}} V (\boldsymbol{x}) \leq 0,
	\end{equation}
	$\forall \boldsymbol{x} \in \mathcal{M} \backslash \{\boldsymbol{x}_*\}$. 
\end{remark}

\begin{figure}
	\centering
	\makebox[0.5\textwidth]{\includegraphics[width=0.66\textwidth]{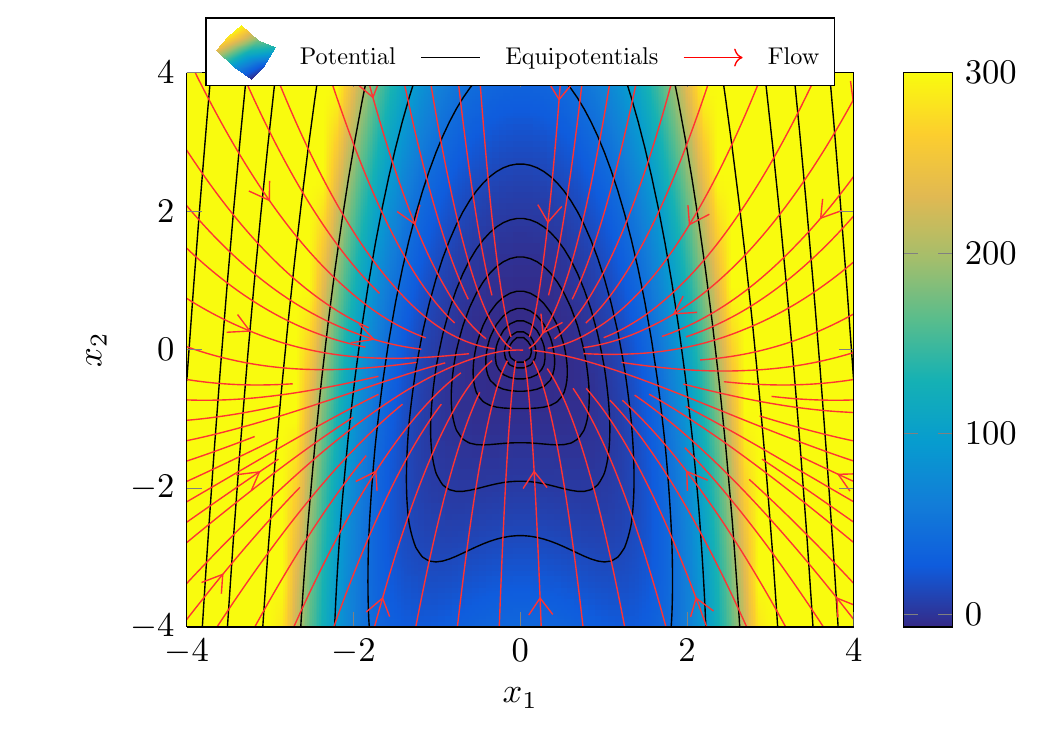}}
	\caption{The Lyapunov function for the nonlinear system from Example \ref{ex:motiv} ($a=0.9$, $b=0.8$) as a sum of squared eigenfunctions $V(\boldsymbol{x})=\boldsymbol{\phi}^{\top}(\boldsymbol{x})\boldsymbol{P}\boldsymbol{\phi}(\boldsymbol{x})$ weighted by the solution of the Lyapunov equation $\boldsymbol{\Lambda}^{\top} \boldsymbol{P} \boldsymbol{\Lambda}-\boldsymbol{P}+\boldsymbol{I}=\boldsymbol{0}$.}
	\label{ogNLSysLyap}
\end{figure}

{The intricate connection of operator spectra and Lyapunov stability for the continuous-time case has not been rigorously extended to discrete-time maps. The transferability of continuous-time results is mentioned in \cite{Mauroy2016b,Mauroy2020} - pointing to the fact that operator-theoretic considerations have the ability to bridge the gap between continuous- and discrete-time settings via the Spectral Mapping Theorem \cite{Pazy1983}. This is a promising proposition, as conventional Lyapunov stability results for nonlinear systems are not directly transferable from one setting to the other. In Figure \ref{ogNLSysLyap} we demonstrate that the Koopman operator spectral analysis of nonlinear maps mirrors classic linear stability analysis, by utilizing linear techniques to deliver a Lyapunov function faithfully representing the nonlinear dynamic flow.}

\subsection{Analysis via the Koopman paradigm}

The Koopman operator paradigm gives a proxy to spectral analysis of a nonlinear system through the eigenfunctions and eigenvalues of the corresponding Koopman operator.
Koopman operators are utilized in the global analysis of complex dynamical systems as their eigenfunctions can be used for a many notions relevant for analysis: metastable sets, splitting the dynamics into dominant slow/fast processes, or to separate superimposed signals \cite{Champion2019, Klus2020a}.
More specifically, Koopman eigenfunctions provide a lot of information about the original system, including a characterization of invariant sets such as stable and unstable manifolds \cite{Brunton2016a}. 
Furthermore, the Koopman paradigm is providing intrinsic information about state space geometry i.e. through level sets of the state space with the same rate of convergence to the attractor - \emph{isostables} and sets with an identical phase value - \emph{isochrons}.
Apart from relating to Lyapunov functions, Koopman eigenfunctinos are used to derive contracting metrics in \cite{Mauroy2013}. {Moreover, equivalences between contraction theory and Koopman operators are explored in \cite{Yi2021}.}
While for dissipative systems the Koopman operator eigenfunctions capture isochrons and isostables, for ergodic systems they capture periodic partitions of the state space \cite{Mezic2005}.
Furthermore, Koopman eigenfunctions with eigenvalue is $0/1$ (continuous/discrete-time) remain constant over trajectories so that its level sets partition the state space into invariant regions - defining the conserved quantity of the system.
Also, the stable unstable and center manifolds of a system can be expressed via the joint level sets of (generalized) Koopman operator eigenfunctions \cite{Mezic2015}.
\begin{proposition}[\cite{Mauroy2013b}]\label{gasC1}
	Consider $\mathbb{X} \subseteq \mathbb{R}^n$ and a vector field $\boldsymbol{f}: \mathbb{X} \mapsto \mathbb{X}$ with a full rank Jacobian $\boldsymbol{f}^{\prime}(\boldsymbol{x}_*)$ and eigenvalues $\operatorname{Re}\{s_i\} < 0 $ at the fixed point $\boldsymbol{x}_*$. If the eigenfunctions of the Koopman operator are $C^1$, then $\dot{\boldsymbol{ x}}=\boldsymbol{f}(\boldsymbol{ x})$ is globally asymptotically stable (GAS) in $\mathbb{X}$.
\end{proposition}
The above result is a global equivalent of the well-known local stability result via Lyapunov's indirect method \cite{khalil1992nonlinear}. Given local stability is implied by stable $\boldsymbol{f}^{\prime}(\boldsymbol{x}_*)$, but global stability is implied by $C^1$-eigenfunctions of the Koopman operator.
Furthermore, works of Mauroy et al. \cite{Mauroy2016b,Mauroy2018,Mauroy2020} establish that joint-level sets of Koopman (generalized) eigenfunctions (for Koopman eigenfunctions from Theorem \ref{gasC1}) offer principled parameterization of candidate Lyapunov functions i.e.
\begin{equation}
{V}(\boldsymbol{x})=\left(\sum_{i=1}^{N}\left|\phi_{{i}}(\boldsymbol{x})\right|^{p}\right)^{\frac{1}{p}},
\end{equation}
for $p \in \mathbb{N}$. As the eigenfunctions can serve as evolution coordinates of nonlinear systems defined by their corresponding eigenvalues, they are directly related the system's contraction metrics. Nevertheless, the Koopman operator paradigm provides a more general stability result than the one of Property \ref{gasC1}. 

\begin{theorem}[\cite{Mauroy2016b}]\label{ggas}
	Let $\mathbb{X}$ be a positive invariant compact set and that the Koopman operator $\mathcal{K}^{t} \phi = \phi \circ \boldsymbol{{F}}^t$ admits an eigenfunction $\phi_s \in C^{0}(\mathbb{X})$ with the eigenvalue $\operatorname{Re}\{s\}<0$. Then the zero level set
	$$
	\mathbb{S}_{0}=\left\{x \in \mathbb{X} \mid \phi_{s}(\boldsymbol{x})=0\right\}
	$$
	is forward invariant under $\boldsymbol{{F}}^t$ and globally asymptotically stable.
\end{theorem}
{The results of Theorem \ref{ggas} are easily recognized in Figure \ref{EFsliceEVO} for the system from Example \ref{ex:motiv}.}

\begin{example}[\cite{Mauroy2016b}]
	To highlight the globality of the spectral analysis via the Koopman operator, consider a non-hyperbolic system $\dot{x}=-x^{3}$ where the analysis via the indirect Lyapunov method is not possible. The system admits a continuous (generalized) eigenfunction with $\phi_{s}(x)=\operatorname{exp} \left(-1 /\left(2 x^{2}\right)\right)$ with the associated eigenvalue $s=-1$. Given that, one can imply global asymptotic stability of the origin by Theorem \ref{ggas}.
\end{example}

\subsection{Other properties via the Koopman paradigm}\label{syspropKoop}
Here we consider some structural properties that can be recognized via the Koopman paradigm's spectral analysis.
It is important to note that the choice of the observables' space conditions the spectral properties of the Koopman operator. The original theory of B.O. Koopman \cite{Koopman1931,Koopman1932} studied the ergodicity of Hamiltonian flows naturally associated with the space of square-integrable observables. 

\begin{remark}[\cite{Lasota1994}]\label{HamKoop}
	For conservative (Hamiltonian) systems the Koopman operator $\mathcal{K}^t: \mathcal{F} \mapsto \mathcal{F}$ is unitary on the Hilbert space $L^2(\mu)$ of complex valued square-integrable functions tied with any Borel probability measure $\mu$ 
	\begin{equation}
	\mathcal{F}=L^{2}(\mathcal{M}), \quad \mathcal{F}=\mathcal{F}^{\dagger},
	\end{equation}
	holding $\forall s$ eigenvalues from the \emph{spectrum}  $\sigma(\mathcal{ K })$ of  $\mathcal{ K }$. Thus, the Koopman operator ${\mathcal{K}}^{t}$ is unitary with $\left|\operatorname{exp}({s})\right|=1$, $s \in \mathbb{C}$. 
\end{remark}
	If the system is measure preserving it possesses an invertible flow $\boldsymbol{F}^t: \mathbb{X} \mapsto \mathbb{X}$. Then the relation between the Koopman operator $\mathcal{K}^t$ and its dual $\mathcal{P}^t$ from Property \ref{dualKP} simplifies to
	\begin{equation}
	\mathcal{K}^t \psi(\boldsymbol{x})=\psi(\boldsymbol{F}^t(\boldsymbol{x})), \quad \quad  \mathcal{P}^t \rho(\boldsymbol{x})=\rho\left(\boldsymbol{F}^{-t}(\boldsymbol{x})\right)
	\end{equation}
	with $\boldsymbol{F}^{-t}=(\boldsymbol{F}^{t})^{-1}$ making the Koopman operator the inverse of its dual.

Dynamical systems with attractor(s) are not energy preserving and thus dissipative, making the Koopman operator non-unitary. 
Furthermore, it is typically non-normal, and can have generalized eigenfunctions \cite{Mezic2019a}.
Dissipative systems are very amendable to the global stability analysis via the Koopman paradigm \cite{Mauroy2016b}.
The choice of the observables-space as $\mathcal{F}=C^{1}(\mathcal{M})$ more easily uncovers \emph{principle eigenfunctions} that are crucial for stability analysis of the equilibrium of dynamical systems \cite{Mauroy2020}. 
The Koopman operators for dissipative systems assemble a contraction semigroup that has a corresponding dissipative generator \cite{ElmarPlischke2005}.
We can redefine the introduced \emph{eigenfunctions} as ordinary eigenfunctions of a linear operator $\mathcal{ K }$ acting in a compact Hilbert space that solve the following
\begin{equation}\label{oEF} 
{\mathcal{K}} \phi_g=\lambda \phi_g \quad \Leftrightarrow \quad ({\mathcal{K}}-\lambda_g) \phi_g = 0 .
\end{equation}
In general, due to the geometric eigenspace being lower than the algebraic eigenspace dimension, a complete basis for the space on which the linear operator acts upon can not always be obtained from (\ref{oEF}), but from \emph{generalized eigenfunctions}.
The notion of \emph{generalized eigenfunctions} in the field is introduced and proved through the case of finite-dimensional linear operators \cite{Mohr2014a,Mezic2019} using Kato decomposition \cite{Kato1966}. In the following we just informally remark on these spectral objects.
\begin{remark}[Generalized eigenfunctions]\label{genEF}
	Generalized eigenfunctions $\xi_g$ are the solution following equality 
	\begin{equation}
	({\mathcal{K}}-\lambda_g)^{m_g} \xi_g = 0
	\end{equation}
	with integers $g=1, \ldots, s$ and $i=0,...,m_g-1$ being the counter of different eigenvalues (geometric eigenspace dimension) and dimensions of algebraic corresponding eigenspaces, respectively.
	Let
	\begin{equation}
	\boldsymbol{\xi}=\left[\xi_{1}^{1}, \ldots, \xi_{1}^{m_{1}}, \xi_{2}^{1}, \ldots, \xi_{2}^{m_{2}}, \ldots, \xi_{s}^{1}, \ldots, \xi_{s}^{m_{s}}\right]^{\top}
	\end{equation}
	be the concatenation of all of the (generalized) eigenfunctions corresponding to the operator.
	Now considering the Koopman operator generator we have
	\begin{equation}\label{geGEF}
	\mathcal{G}_{\mathcal{ K }}{\xi}_{g}^{i}=\dot{\xi}_{g}^{i}=\lambda_{g} \xi_{g}^{i}+\xi_{g}^{i+1},
	\end{equation}
	that in the finite-dimensional case gives 
	\begin{equation}
	\dot{\boldsymbol{\xi}}=\operatorname{diag}\{\boldsymbol{J}_{\lambda_{1}}, \dots, \boldsymbol{J}_{\lambda_{s}}\}\boldsymbol{\xi},
	\end{equation}
	where $\boldsymbol{J}_{\lambda_g}$ is the familiar Jordan block (\ref{JB}).
\end{remark}
\begin{example}\label{ex:gefs}
	Consider the case of a multiplicity of $m_g=2$ in (\ref{geGEF}) resulting in $\dot{\xi}_{g}^{1}=\lambda_{g} \xi_{g}^{1}+\xi_{g}^{2}$. The function $\xi_{g}^{2}$ in turn represents the one ordinary eigenfunction $\phi_g$ satisfying $\dot{\phi}_{g}=\lambda_{g} \phi_{g}$ giving the relation $(\mathcal{G}_{\mathcal{ K }}-\lambda) \xi_{g}=\phi_g$.
\end{example}

\begin{remark}\label{fpSYS}
	The utility of \emph{generalized eigenfunctions} in spanning Koopman-invariant subspaces arising from (\ref{genEF}) generalizes to infinite-dimensional case for discrete spectrum of the Koopman operator. This is courtesy of a \emph{compact} operator defined on a Hilbert space (as it is a limit of finite-rank operators) \cite{Conway2007}.
\end{remark}

\section{Koopman-based control approaches and applications}\label{secV}

Here, the focus is turned towards non-autonomous systems and approaches that still allow for exploiting the Koopman operator paradigm to find efficient representations of control systems.

\subsection{Extensions to control systems}
As the Koopman operator dynamical models are naturally formed for autonomous systems, they require modifications in the non-autonomous case. We present structured ways of including control for classes of nonlinear systems and relate to the core of the Koopman operator theory. In order to do so we take an ``atomic'' approach based on the Koopman eigenfunction evolution PDE (Theorem \ref{charKG}). 

\subsubsection{Control-affine nonlinear systems}
Here we offer a novel derivation of Koopman operator eigenfunctions in the case of control-affine systems. Let us consider the class of such systems, in a single-input and single-output (SISO) form  
\begin{equation}\label{ctrlAff}
\begin{aligned}
\dot{\boldsymbol{x}} & =\boldsymbol{f}(\boldsymbol{x})+\boldsymbol{g}(\boldsymbol{x}){u}  \\
{y} & ={h}(\boldsymbol{x}),
\end{aligned}
\end{equation} 
with $\boldsymbol{f}, \boldsymbol{g} \in \mathcal{M}$ the \emph{drift} and \emph{control} vector fields, respectively. By looking at the time-evolution of the output/observable $y$
\begin{equation}\label{lieEQ}
\dot{y} = \mathcal{L}_{\boldsymbol{f}} h(\boldsymbol{x})+ \mathcal{L}_{{\boldsymbol{g}}} h({\boldsymbol{x}})u ,
\end{equation}
one obtains the familiar Lie-derivative expression. To have a globally linear description of such a system, one can study the eigenspaces of the Lie-derivate operators (Koopman operator generators) acting on the drift and control vector fields.

\begin{theorem}[Bilinear eigenfunctions]\label{thm:Bili}
	Consider the control affine system (\ref{ctrlAff}) with the eigenfunctions $\phi_i$ and $\phi_{c,i}$ of the drift and control vector fields, respectively.
	If $\operatorname{span}\{\phi_{c,1},\phi_{c,2},\dots\} \subseteq \operatorname{span}\{\phi_1,\phi_2,\dots\}$ the eigenfunction time-evolution is bilinearizable.
	
	\begin{proof} 
		Given that $\operatorname{span}\{\phi_{c,1},\phi_{c,2},\dots\} \subseteq \operatorname{span}\{\phi_1,\phi_2,\dots\}$, we can write  ${\phi}_{c}(\cdot)= {\mathcal{C}} {\phi}(\cdot)=\sum_{i=1}^{\infty} {c}_i{\phi}_i(\cdot)$. Then, the eigenfunction evolution for (\ref{lieEQ}) results in
		\begin{equation}\label{biliExp}
		\begin{aligned}
		\dot{{\phi}}(\boldsymbol{x}) &= \mathcal{L}_{{\boldsymbol{f}}} {\phi}(\boldsymbol{x})+ \mathcal{L}_{{\boldsymbol{g}}} \phi_c({\boldsymbol{x}})u \\
		&= (\mathcal{L}_{{\boldsymbol{f}}} + \mathcal{L}_{{\boldsymbol{g}}} {\mathcal{C}}u ) \phi({\boldsymbol{x}}).
		\end{aligned}
		\end{equation}
		As $\mathcal{L}_{{\boldsymbol{g}}}$ and ${\mathcal{C}}$ linear operators their composition $\mathcal{L}_{{\boldsymbol{g}}} {\mathcal{C}}$ is as well. Thus, the eigenfunction evolution becomes bilinearizable.
	\end{proof}
\end{theorem}

\begin{corollary}
	By integrating the infinitesimal expression (\ref{biliExp}) we obtain
	\begin{equation}\label{contCA}
	[{\mathcal{K}}^{t}(u)\phi](\boldsymbol{x}) =\operatorname{exp}[(\mathcal{L}_{{\boldsymbol{f}}} + \mathcal{L}_{{\boldsymbol{g}}} {\mathcal{C}}u ) t] \phi({\boldsymbol{x}}),
	\end{equation}
	giving rise to a control-parameterized family of time-$t$ semigroups of Koopman operators $\{{\mathcal{K}}^{t}(u)\}_{t \in \mathbb{R}_{0,+}}$. Thus, the following ensues for the discrete-time case 
	\begin{equation}\label{Klpv}
	[{\mathcal{K}}(u)\phi](\boldsymbol{x}) =\mathcal{K}_{\boldsymbol{f}} (\mathcal{K}_{\boldsymbol{g}}\mathcal{C})^u \phi({\boldsymbol{x}}),
	\end{equation}
	following trivially by fixing the sampling time $t=t_s$ in (\ref{contCA}). 
\end{corollary}
One can see how the bilinear system theory of the continuous-time case (\ref{contCA}) can now be replaced by linear parameter-varying (LPV) theory \cite{Mohammadpour2012} in the discrete-time case. 
Generalizing (\ref{Klpv}) for multiple-input and multiple-output form is straight forward.

\begin{remark}[State-independent vector field]
	For simpler control-affine systems with state-independent control vector fields i.e. $\dot{\boldsymbol{x}}=\boldsymbol{f}(\boldsymbol{x})+\boldsymbol{b} {u}$, any drift vector field  admits the trivial principle eigenfunction $\phi(\cdot)=1$ with eigenvalue $0$ allowing a bilinear representation without needing the conditions from Theorem \ref{thm:Bili}. Another less restrictive option is to rewrite (\ref{biliExp}) by the chain rule \cite{Kaiser2017a}
	\begin{equation}
	\begin{aligned}
	\dot{\phi}(\boldsymbol{x})&= s \phi(\boldsymbol{x})+{\frac{\partial}{\partial \boldsymbol{x}}  \phi(\boldsymbol{x}) \boldsymbol{B}} \boldsymbol{u},
	\end{aligned}
	\end{equation}
	with $s$ the eigenvalue corresponding to $\phi$, but leads to a state-dependent input term in the evolution of eigenfunctions. 
\end{remark}

\begin{example}[Koopman-bilinearizable system]\label{ex:ctrlBili}
	Consider a continuous-time control system inspired by \cite{Goswami2018}:
	\begin{equation}
	\dot{\boldsymbol{x}} = \left[\begin{array}{c}
	c x_{1} \\
	d x_{2}+\left(d-c^{2}\right) x_{1}^{2}
	\end{array}\right] + \left[\begin{array}{cc}
	\boldsymbol{g}_1(\boldsymbol{ x}) & \boldsymbol{g}_2(\boldsymbol{ x})
	\end{array}\right]\boldsymbol{u},
	\end{equation}
	with control vector fields $\boldsymbol{g}_1(\boldsymbol{ x})=[1, x_{1}^{2}]^{\top}$. $\boldsymbol{g}_2(\boldsymbol{ x})=[0, 1]^{\top}$
	The system from can be \emph{lifted} into eigenfunction-coordinates $\boldsymbol{\phi}(\boldsymbol{x})=[x_1, x_2+ x_1^2, x_1^2, 1]^{\top}$ of the Koopman generator with the diagonal transition matrix of respective eigenvalues $\boldsymbol{{ \Lambda }}=\operatorname{diag}\{c, d, c^2, 0\}$. Then, we can write
	
	\begin{equation}	
	\dot{\boldsymbol{z}} = \boldsymbol{A} \boldsymbol{z} + \boldsymbol{B}_{1} \boldsymbol{z} u_{1}+\boldsymbol{B}_{2} \boldsymbol{z} u_{2},
	\end{equation}
	by lifting the state $\boldsymbol{x}$ to $\boldsymbol{z}=\boldsymbol{\phi}(\boldsymbol{x})$.
	After applying (element-wise) the control vector field Lie-derivatives with respect to the new coordinates, we obtain
	\begin{equation}
	\boldsymbol{B}_{1}=\left[\begin{array}{cccc}
	0 & 0 & 0 & 1 \\
	2 & 0 & 1 & 0 \\
	2 & 0 & 0 & 0 \\
	0 & 0 & 0 & 0
	\end{array}\right], \boldsymbol{B}_{2}=\left[\begin{array}{cccc}
	0 & 0 & 0 & 0 \\
	0 & 0 & 0 & 1 \\
	0 & 0 & 0 & 0 \\
	0 & 0 & 0 & 0
	\end{array}\right],
	\end{equation}
	allowing us to write $\mathcal{L}_{\boldsymbol{g}_i}\boldsymbol{\phi}(\boldsymbol{ x})=\boldsymbol{B}_i\boldsymbol{z}$.
\end{example}

Although the the eigenfunction evolution is bilinear, there are well studied tools to deal with such problems cf. \cite{Elliott2009}.

\begin{remark}[Representation heuristic]
	Works considering Koopman operator dynamical modeling often assume the following evolution $\dot{\boldsymbol{z}}= \boldsymbol{A} \boldsymbol{z}+\boldsymbol{B} \boldsymbol{u}$ in the "lifted" $\boldsymbol{z}$-coordinates. Clearly, such a representation of the control effect is only local. Thus, the results of this section give a theoretical justification on why many data-driven Koopman operator-related approaches for control systems are only well-versed for short-term prediction.
\end{remark}

\subsubsection{General nonlinear control systems}
As opposed to control-affine systems, general nonlinear control systems of the form $\dot{\boldsymbol{x}}=\boldsymbol{f}(\boldsymbol{x},\boldsymbol{u})$ do not admit a closed form representation for (bi)linear prediction via the Koopman operator paradigm. Nonetheless, meaningful representations are still possible. One can assume the existence of joint state-and-input eigenfunctions \cite{Kaiser2017a}, leading to the following eigenfunction PDE
\begin{equation}
\begin{aligned}
\dot{\phi}(\boldsymbol{x},{\boldsymbol{u}})& =\mathcal{L}_{{\boldsymbol{f}}} \phi(\boldsymbol{x},{\boldsymbol{u}})+\frac{\partial}{\partial \boldsymbol{u}}\phi(\boldsymbol{x},{\boldsymbol{u}}) \dot{\boldsymbol{u}},
\end{aligned}
\end{equation}
where $s$ is the eigenvalue corresponding to $\phi$.
Given the relation above, the derivative of control $\dot{\boldsymbol{u}}$ is the input to eigenfunction coordinates leading to integral or incremental control in the discrete-time analogue. Nevertheless, the form admits general nonlinear control systems. 
In a similar vein, Korda et al. \cite{Korda2020b} make the case that the form is non-restrictive as any system can be transformed by ``state inflation'' $\dot{\boldsymbol{v}}=\boldsymbol{u}$ via $\boldsymbol{x}=[\boldsymbol{\zeta}^{\top}, \boldsymbol{v}^{\top}]^{\top}$ into
\begin{equation}
\underbrace{\left[\begin{array}{c}
	\dot{\boldsymbol{\zeta}} \\
	\dot{\boldsymbol{v}}
	\end{array}\right]}_{\dot{\boldsymbol{x}}}=\underbrace{\left[\begin{array}{c}
	{\boldsymbol{f}}_{\boldsymbol{\zeta}}(\boldsymbol{\zeta},\boldsymbol{v}) \\
	\boldsymbol{0}
	\end{array}\right]}_{\boldsymbol{f}(\boldsymbol{x})}+\underbrace{\left[\begin{array}{l}
	\boldsymbol{0} \\
	\boldsymbol{I}
	\end{array}\right]\dot{\boldsymbol{v}}}_{\boldsymbol{B}\boldsymbol{u}},
\end{equation}
again limiting our direct control to controlling the rate of change of the input signal.
As a single Koopman generator corresponds to a single ODE or PDE, the introduction of input dependence parameterizes the differential equations with a possibly infinite set of inputs. However, restricting those to a finite set can be represented as a system of switched Koopman operators. The authors in \cite{Peitz2019} utilize this for efficient model predictive control. 
{
	\subsection{Control-oriented frameworks}
	Here we survey notable options for designing control-oriented frameworks based on the Koopman operator theory (for a graphical overview cf. Figure \ref{fig:graphCTRL}).
	
		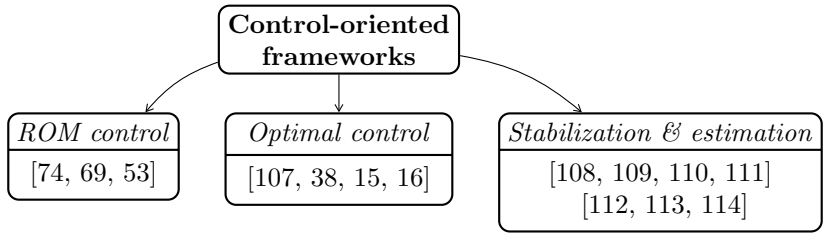
\begin{figure*}[ht!]
		\centering
		
		\begin{tikzpicture}[
		node distance = 5mm and 5mm, 
		box/.style = {draw, rounded corners, 
			minimum width=15mm, minimum height=7mm, align=center,draw=black, thick,},
		> = {Straight Barb[angle=60:3pt ]},
		bend angle = 15,
		auto = right,
		state/.style={
			rectangle split,
			rectangle split parts=2,
			rectangle split part fill={red!0,blue!0},
			rounded corners,
			draw=black, thick,
			minimum height=30mm,
			minimum width=22mm,
			inner sep=3pt,
			text centered,
		}
		]
		
		\node (n3)  [box]  
		{\textbf{Control-oriented} \\ \textbf{frameworks}};
		\node (n7)  [state, below =of n3,fill=blue!0]    
		{\textit{Optimal control}  \nodepart{two}   $\begin{array}{c}
			\text{\cite{Korda2018a,Folkestad2019,Korda2020b,Igarashi2020}}
			\end{array}$ };
		\node (n5)  [state, below right =of n3,fill=red!0] 
		{\textit{Stabilization \& estimation} \nodepart{two}   $\begin{array}{c}
			\text{\cite{Das2019,Ma2019,Huang2019a,Narasingam2020}} \\\text{ \cite{Surana2016,Surana2016a,Netto2018}}
			\end{array}$ };
		\node (n6)  [state, below left =of n3,fill=blue!0]   
		{\textit{ROM control}  \nodepart{two}  $\begin{array}{c}
			\text{\cite{Kaiser2019,Peitz2019,Klus2020}}\end{array}$ };
		
		\draw[->] (n3) to [bend left,""]  (n5);
		\draw[->] (n3) to [bend right,""]  (n6);
		\draw[->] (n3) to [""]  (n7);
		
		\end{tikzpicture}
		\caption{Main branches of control-oriented approaches using the Koopman operator-based models. The bottom split part of the latter includes relevant references employing the approach noted in the top part.}
		\label{fig:graphCTRL}
	\end{figure*}

	\subsubsection{Reduced order model (ROM) control}
	One simplifying perspective looks to use parts of the intrinsic dynamics information from the Koopman operator for energy-based control or reduced order controllers. These approaches utilize Property \ref{HamKoop} for identifying the Hamiltonian from data - as it corresponds to a Koopman eigenfunction \cite{Kaiser2019}. 
	Furthermore, when it comes to reduced order modeling the Koopman operator paradigm is put to use in \cite{Peitz2019,Klus2020} for data-driven switched control of systems governed by partial differential equations. 

}
\subsubsection{Optimal control}

For optimization-based approaches such as model predictive control (MPC), the evident benefit is the increased accuracy and validity in contrast to linear models based on a local linearization, while having a reduced computational burden compared with schemes based on nonlinear models. A comparison of the outputs of a convex program with Koopman-based (and another global linearization) models versus the solutions of the original nonlinear programming problem can be found in Igarashi et al. {\cite{Igarashi2020}}.
For a principled MPC treatment via the Koopman operator paradigm we point to Korda et al. \cite{Korda2018a} with subsequent optimized versions for more efficacious prediction in Korda et al. \cite{Korda2020b} and Folkestadt et al. \cite{Folkestad2019}. 

\subsubsection{Stabilization and state-estimation} 
Regarding stabilization, lifted space-based control Lyapunov functions have been considered in a range of works \cite{Das2019,Ma2019,Huang2019a, Narasingam2020} - showing promise for nonlinear optimal stabilization. 
Moreover, works on observer synthesis \cite{Surana2016,Surana2016a} and filtering \cite{Netto2018} show improved efficacy via the use of Koopman operator dynamical models.

\subsection{Why not linearize via feedback?}
The idea of embedding a system into possibly a higher-dimensional linear system is not a new concept in the control community \cite{Lee1988}.
One of the reasons non-local linear representations of nonlinear dynamics could be appealing is efficient controller design - both regarding computation and control effort. The considered Koopman operator-based linearizations do not require compensating input injections as their exact linearization counterparts such as input-output and feedback linearization. For systems with useful (stabilizing) nonlinearities, feedback linearization approaches are known to be non-robust \cite{Kothare1995}. Furthermore, feedback linearizing controllers does not aid in respecting arbitrary cost functionals (in a receding or infinite horizon fashion) for control design due to compensation-based policies \cite{Freeman1995}. One example of the advantageous use of Koopman operator dynamical models with respect to feedback linearization can be found in Kaiser et al. \cite{Kaiser2017a}. 

An interesting category of dynamically challenging systems are underwater vehicles, e.g. remotely operated vehicles (ROVs). They are nonlinear and stable due to the dampening effect of hydrodynamic forces. Thus, ROVs present a real world example of a system with useful nonlinearities, whose cancellation (i.e. via feedback linearization) can render a stable system unstable for the smallest of modeling errors \cite{Freeman1996} while using excessive amounts of control effort \cite{Krstic1998}. With the recent developments in efficient bio-inspired designs of underwater vehicles \cite{7426526,Zhueaax4615}, the need for more efficacious controllers becomes increasingly important for their autonomous operation. 
Given that, it is worthwhile considering Koopman operator dynamical models for efficient optimal control, especially for ``well-behaved" systems.

\section{CONCLUSION}\label{secVII}

We systematically review all key aspects of \emph{Koopman operator dynamical models} in order to provide information relevant for data-driven representations, analysis and control through a holistic perspective.
Concerning representations, the current state-of-the-art looks to be geared more towards finding relevant eigenfunctions for prediction, as opposed to a span of only approximately Koopman-invariant observables without tapping into the operator's spectrum directly. Current approaches are able to well represent a system's transient and quasi-periodic behavior while non-dissipative dynamics and continuous spectra are still an open research question. The compelling feature of the paradigm is that, if representations can be efficiently learned in a well-posed manner, a system's prediction, analysis and control become greatly simplified. At present, that involves employing modeling approaches for which there is no overarching data-driven framework. Nevertheless, the recent developments in the field show promise in making Koopman operator-based system modeling more viable than ever before.  	
When the operator's eigenfunctions are shared between the continuous- and discrete-time representations of a system, they form meaningful and amendable parameterizations of the system's dynamics (even if learned via non-parametric methods).

The one main challenge in the field of Koopman operator dynamical models is learning genuinely invariant coordinates under the Koopman operator.
Moreover, the relevance determination of the coordinates together with model order choice require a more rigorous consideration.
In terms of safety-critical control, tractable \emph{a priori} error quantification is still unconquered - hindering the wider applicability of Koopman operator-based frameworks. 
Also, the linear evolution has the potential to offer efficient uncertainty propagation - a more challenging matter for models in ``dynamics of states" representations.

{While the heuristics of current data-driven approaches do work well under some circumstances, there is evidence pointing to the fact that \textit{spectrally direct} approaches lead to better performance with more efficacious lifting bases. Nevertheless, there is still a considerable gap between the Koopman operator-based system-theoretic considerations and their data-driven realizations - due to the incoherence between the system and learning theory. Future works addressing this, at least for some classes of systems, are of great importance on the way towards the development of an overarching framework. In turn, this would allow meaningful comparisons to conventional system identification and control methods - which is currently not possible in a principled way. Moreover, rigorous results on the general existence of finite-dimensional exact representations, akin to the one from the motivating Example \ref{ex:motiv} are missing. Also, rigorous results on spectral convergence and the convergence of modes (to reconstruct the observable to predict) are still an open question. This analysis would split genuine time-linearity of the considered coordinates and their relevance when one can meaningfully truncate the spectral decomposition. \textit{A priori} certificates regarding the quality of state-reconstruction given a desired linear system order - as well as the converse - are still not rigorously considered. Nevertheless, the modes and the spectrum are interconnected, but if the learner is constrained to only consider the Koopman-invariant coordinates - i.e. (generalized) eigenfunctions - one could focus on finding a subset of those that efficiently reconstruct the observable to predict.
}

Furthermore, the general lack of comparison of the resulting control frameworks to exact linearization procedures using feedback needs to be addressed. Such comparisons could better rectify under what circumstances the Koopman operator-based controllers are actually advantageous. It is also crucial to assess the robustness of these controllers under external disturbances - something not addressed at present. Nevertheless, the limits of predicting general controlled systems demonstrated in Section \ref{secV} make the consideration of iterative direct control design is more conceivable than obtaining globally predictive models. For model-based reinforcement learning the paradigm poses an interesting prospect, as an almost global linear model implicitly parameterizes both the optimal value function and the policy (through a set of \emph{lifted} coordinates).
Under changing conditions, exploring the possibilities of adaptive control and online learning remain an exciting area yet to be investigated.

\bibliography{KoopSurv}

\end{document}